\documentclass[a4paper, 12pt]{article}  
\usepackage{pdfpages}
\usepackage{lmodern}
\usepackage{graphicx} 
\usepackage{wrapfig} 
\usepackage{textcomp} 
\usepackage[utf8]{inputenc}
\DeclareSymbolFont{letters}{OML}{ztmcm}{m}{it}
\usepackage{color}
\usepackage{circuitikz}
\usepackage{bm}
\usepackage{bbm}
\usepackage{mathtools,tikz,caption}
\usetikzlibrary{decorations.pathmorphing, patterns,shapes}
\DeclareRobustCommand\sampleline[1]{%
  \tikz\draw[#1] (0,0) (0,\the\dimexpr\fontdimen22\textfont2\relax)
  -- (0.99em,\the\dimexpr\fontdimen22\textfont2\relax);%
}
\newcommand{\hide}[1]{}

\newcommand{\beno}{\begin{equation}\begin{array}{llllllllll}\nonumber}
\newcommand{\be}{\begin{equation}\begin{array}{llllllllll}}
\newcommand{\ee}{\end{array}\end{equation}}
\usepackage{multirow}

\usepackage{array}
\usepackage{listings}
\usepackage{adjustbox}
\usepackage[boxed]{algorithm2e}
\usepackage{subcaption}
\usepackage{multirow} 
\usepackage[OMLmathrm,OMLmathbf]{isomath}
\usepackage{booktabs} 
\usepackage{pdflscape}
\usepackage{natbib}
\usepackage{multibib}
\usepackage{multicol}
\newcites{supp}{Supplementary Material References}
\usepackage{rotating}

  \usepackage[T1]{fontenc} 
\usepackage{geometry}
\usepackage{tablefootnote}
\usepackage{amsmath}
\usepackage{etoolbox} 
\usepackage{amsthm}
\usepackage{amsfonts}
\usepackage{amssymb}
\usepackage[section]{placeins}
\usepackage{diagbox} 
\usepackage{hyperref}
\usepackage{graphicx} 
\usepackage{fixmath} 
\usepackage{ulem} 
\usepackage[scaled=.90]{helvet}
\usepackage{setspace} 

\makeatletter
\renewcommand\@biblabel[1]{\textbf{#1.}} 
\usepackage{enumitem}
\newcommand{\blind}{1}
\usepackage{pifont}
\setlist{nolistsep,leftmargin=*}
	\newcommand{\dsum}{\displaystyle\sum\limits}
 
\usepackage{tikz} 
\usetikzlibrary{backgrounds}
\usetikzlibrary{calc}

\definecolor{grey}{rgb}{0.5,0.5,0.5}
\definecolor{lightgrey}{rgb}{0.7,0.7,0.7}
\definecolor{verylightgrey}{rgb}{0.9,0.9,0.9}
\definecolor{darkgrey}{rgb}{0.2,0.2,0.2}

\definecolor{grey5}{rgb}{0.9,0.9,0.9}
\definecolor{grey4}{rgb}{0.833,0.833,0.833}
\definecolor{grey3}{rgb}{0.766,0.766,0.766}
\definecolor{grey2}{rgb}{0.7,0.7,0.7}
\definecolor{grey1}{rgb}{0.533,0.533,0.533}

\usepackage{xcolor}
\usepackage{graphicx}
\usepackage{caption}
\usepackage{tocloft}
\usepackage{etoolbox}
\usepackage{titletoc}
\definecolor{foreigners}{HTML}{FF8787}
\definecolor{chileans}{HTML}{27408B}

\usepackage{dlfltxbcodetips,bbm,mathtools,amsfonts,amsmath,amssymb,fancybox,graphicx,bm,latexsym}
\usepackage[mathscr]{eucal}

\renewcommand{\maketitle}{ 
	\begin{center}
		{\LARGE\@title} 
		
		\vspace{0pt} 
		
		{\large\@author} 
		\\\@date 
		
		\vspace{40pt} 
	\end{center}
}
\date{ \vspace{1ex} \normalsize \today}

\begin{document}
	
	\def\spacingset#1{\renewcommand{\baselinestretch}%
		{#1}\small\normalsize} \spacingset{1}

	
	\if1\blind
	{
	
		\title{\textbf{Socio-cognitive Networks between Researchers: Investigating Scientific Dualities with the Group-Oriented Relational Hyperevent Model \\} \vspace{.1cm}}
		\author{Alejandro Espinosa-Rada$^{1,2}$, J\"urgen Lerner$^3$, Cornelius Fritz$^4$,  
        \hspace{.2cm}\\ $^1$ Social Network Science Group-UC, Department of Sociology, Pontificia Universidad Cat\'olica de Chile
        \hspace{.2cm}\\ $^2$ Social Networks Lab, ETH Zurich
        \hspace{.2cm}\\ $^3$ University of Konstanz, Germany
        \hspace{.2cm}\\ $^4$ School of Computer Science and Statistics, Trinity College Dublin\\ ~ }
		\maketitle
	} \fi

	\if0\blind
	{
		\bigskip
		\bigskip
		\bigskip
		\begin{center}
			{\LARGE\bf Socio-cognitive Networks between Researchers: Investigating Scientific Dualities with the Group-Oriented Relational Hyperevent Model}
		\end{center}
		\medskip
	} \fi
	
	\bigskip
\begin{abstract}
    Understanding why researchers cite certain works remains a key question in the study of scientific networks. Prior research has identified factors such as relevance, group cohesion, and source crediting. However, the interplay between cognitive and social dimensions in citation behavior—often conceptualized as a socio-cognitive network—is frequently overlooked, particularly regarding the intermediary steps that lead to a citation. Since a citation first requires a work to be published by a set of authors, we examine how the structure of coauthorship networks influences citation patterns. To investigate this relationship, we analyze the citation and collaboration behavior of Chilean astronomers from 2013 to 2015 using the Group-Oriented Relational Hyperevent Model, which allows us to study coauthorship and citation networks in a joint framework. Our findings suggest that when selecting which works to cite, authors favor recent research and maintain cognitive continuity across cited works. At the same time, we observe that coherent groups—closely connected coauthors—tend to be co-cited more frequently in subsequent publications, reinforcing the interdependence of collaboration and citation networks.
    \noindent

{\em Keywords:} Duality, Scientific Networks, Science of Science, Relational Event Model, Hyperevents.

\end{abstract}

\spacingset{1.5} 

\hide{
\paragraph*{Highlights}
\begin{itemize}
    \item Joint analysis of social and cognitive dimensions enables a more realistic model of scientific networks.
    \item Proposal of a novel Group-Oriented Relational Hyperevent Model for analyzing group-to-set events.
    \item When selecting works to cite, authors favor recent publications and maintain cognitive continuity between cited works.
    \item Coherent coauthor groups tend to be co-cited more frequently, underscoring the link between collaboration and citation.
\end{itemize}

\paragraph*{Funding}

The authors acknowledge support from  Fondecyt Regular (ANID) under grant number 1220560 (AER) and the German Research Foundation under grants 321869138 (JL) and 520770522 (CF).

\paragraph*{Data availability statement}

For legal reasons, data from Clarivate Web of Science cannot be made openly available.
}

\section{Introduction}
\label{sec:introduction}

Why do researchers cite each other? Citation is one of the most relevant indicators to measure the history of knowledge, and the impact and recognition of researchers, which, as a consequence, reinforces some level of hierarchy in science \citep{Price_1965, Crane1972, Cole_Cole_1973, Merton_1988, Bellotti_Espinosa}. Still, understanding why researchers cite each other is a longstanding conjecture in studying scientific networks. Existing theories suggest that authors cite relevant contributions, cite their group of reference, or cite other work because of its honest contribution to their publications \citep{Nicolaisen_2007}. \citet{White2004} also emphasize interpersonal networks, such as coauthorship networks. However, prior research often overlooked the dual nature of cognitive and social dimensions \citep{Mutzel2020}.

Citations help researchers trace existing knowledge, understand its diffusion, and examine how peer recognition shapes intellectual contributions. Knowledge evolves through recursive framing, influenced by social structures such as communication, interactions, and relations among scholars, mentorship, and groups. This interplay between knowledge and social structures forms a socio-cognitive network, where thinking itself is shaped by social interactions and the knowledge available. In citation practices, references reflect both intrinsic intellectual contributions and the influence of scholarly networks. The concept of dualities \citep{Breiger_1974} allows researchers to analyze these intertwined processes, highlighting the mutual impact of groups of scholars and knowledge development.

While different interpretations exist in the literature, there is no agreed-upon theory on how social or cognitive ties are interrelated and which are the main mechanisms driving accumulation processes. 
Are citations accumulated because of intellectual merits or because individuals know each other? To unravel this puzzle, we investigate the tendency of researchers' citations by examining the concurrent interplay of socio-cognitive ties through coauthorship and citation networks. 

Understanding citation patterns is challenging, although they have been extensively criticized as a simplistic measure \citep{Edge_1979} that is often under-theorized \citep{Leydesdorff1998, Nicolaisen_2007}. However, to disentangle their structural properties, recent research has analyzed how they are embedded in a social context to identify the main mechanisms underlying why authors cite each other and how different networks are interrelated to explain citation tendencies \citep{White2004, espinosa2024, lerner2024relational_coevolution}. This literature followed the tradition of the network researchers working on the sociology of science and knowledge that investigates social circles \citep{Bellotti_Espinosa}, as groups that are comprised of scientists who work on similar research problems that are usually aware of each other and maintain a high level of informal communication allowing them to navigate the complexity of science by creating social organizations of this kind beyond their institutional affiliations \citep{Crane_1969}. These social circles or invisible colleges can be investigated using bibliographic data through the lens of the duality of socio-cognitive ties \citep{Kadushin_1966, Crane_1969, Breiger_1974}.

\begin{figure}[t!]
    \centering
    	\includegraphics[width = .48\linewidth, keepaspectratio]{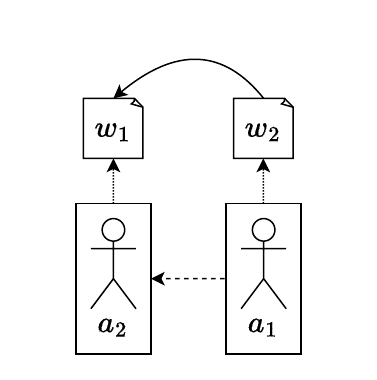}
\caption{Illustrative example of indirect and direct interactions between actors and publications.
}
\label{fig:example}

\end{figure}

The literature on author citation networks often relies on aggregate data and has not yet explored the intermediary role of works without aggregating the information (e.g., \citealp{Small_1973} and 
\citealp{Zhao_Strotmann_2008}). This gap arises due to difficulties considering the complex and interdependent latent mechanisms in citation data. As illustrated in Fig. \ref{fig:example}, for actor $a_1$ to cite actor $a_2$, $a_1$ must first publish work\footnote{Note that we use the terms publication and work interchangeably.} $w_2$ referencing another work $w_1$ (co-)authored by actor $a_2$. The numerous ways of representing the citation process underscore the socio-cognitive dimension of scientific networks and give rise to multiple interdependent \textsl{``dualities''}. 
Authors become indirectly related by coauthoring the same work, citing the same works or authors, and being cited by the same works or authors. The authors become indirectly related via common authors – but also by citing, or being cited by, the same works or authors. Considering multiple paths reveals various \textsl{``dualities''} that depend on one another. 
Fig. \ref{fig:different_representation} further illustrates these dualities.  
For instance, the citation relationship between two works may depend on the coauthorship relations between the authors. Disregarding this dependence results in information loss and biased results. In what follows, nodes correspond to scholars and works, and ties to (co)authorship and citation.

In this paper, we explore the dual relationship between academic citations and coauthorship to shed light on this complex phenomenon by focusing on how authors cite other authors through works.
We start by laying out the available theory on the concept of \textsl{``duality''} and bibliographic patterns in Section \ref{sec:theory}. 
In Section \ref{sec:socio-cognitive}, we theorize about academic authors' interdependencies between citation and collaboration behavior to identify whether the collaboration begets citations and more collaborations. 
As a case study, we analyze the collaboration and citation data among Chilean astronomers from 2013 to 2015, introduced in Section \ref{sec:data}. 
We then propose a statistical model for the available data  in Section \ref{sec:methods} unifying previous research findings on citation and coauthor behavior (e.g., Matthew effect and group effect) and 
our novel theoretical insights about the interplay between coauthorship and citation. 
We hypothesize that effects drawing on \textsl{``dualities''} in socio-cognitive networks among scientific works and researchers provide additional explanations for \textsl{``who coauthors with whom''} and \textsl{``who cites whom''.} 
This model allows us to simultaneously consider different \textsl{``dualities''} in scientific networks by studying citations and collaboration ties in a joint framework. Thus, we advance the duality approach by using all available information in bibliometric data.
Against this background, we assess our claims by employing a group-oriented variant of the relational hyperevent model \citep{lerner2023relational_polyadic,lerner2024relational_coevolution} in Section \ref{sec:results}. Finally, Section \ref{sec:discussion} provides a discussion of the main findings.

\section{Theory}
\label{sec:theory}

Citations help researchers identify the knowledge available to others and examine how this knowledge spreads and evolves through peer recognition and scholarly contributions. 
Network researchers argue that the emergence and evolution of ideas -- understood as sets of beliefs and bodies of knowledge -- are produced, in part, by groups of scientists and that it is the social order governing such groups that shapes the way ideas are formed and accepted \citep{Bellotti_Espinosa}. 
Knowledge is framed by existing ideas about what is already known, serving as an initial reference point. 
These frames are constructed from available and necessary information, shaped by how it is gathered and processed, and refined through a recursive process that enables researchers to develop new conceptual frames \citep{carley1986}. 
Groups of scientists further modulate these frames, encompassing regular communication, interactions, and relationships among researchers, participation in ``coherent groups'', intergenerational mentor-mentee networks, and other intellectual settings. 
These structures serve as pathways to an individual's cognitive processes. 
As an individual’s frame evolves and becomes accessible to others, it can influence their position within the social network and the knowledge they engage with, and vice versa \citep{Mullins_Mullins_1973, carley1986, collins2002}.

The intrinsic relationship between knowledge production, dissemination, and underlying social structures is known as a socio-cognitive network, as these elements are fundamentally interconnected \citep{carley1986, White2004}. 
As a cultural practice, knowledge is learned, discussed, and shared among scholars with similar interests. 
Thus, thinking can be perceived as internalizing cognitive structures shaped by available knowledge and social networks. 
As an individual's frame evolves, their cognitive understanding and position relative to others within the network may shift.

In the context of citations, the socio-cognitive network is partly reflected in how references to others' works and their authors diffuse—whether or not the cited authors are personally known to the person citing. 
Behind each act of referencing are authors and research teams that collaborate and collectively decide which works to cite as a foundation for further research. 
These references may be selected based on the intrinsic ideas they convey -- regardless of authorship as publicly available knowledge -- or due to the relevance of the scholars behind them, who function as ``coalitions in the mind'' \citep{collins2002} and serve as intellectual reference groups. 
This interconnectedness underscores the socio-cognitive nature of citation networks. 
Moreover, developments using dualities \citep{Breiger_1974} have enabled researchers to represent these relationships and assess the relative importance of different processes shaping the network.

\textsl{``Duality''}, as a concept, was initially associated with the intersection of social circles following the tradition of Georg Simmel. 
This concept was related to relationships among actors of different levels (e.g., individuals and organizations) through membership relations \citep{Breiger_1974}. 
The main reasons for the intersections are individuals' shared interests, personal affinities, or ascribed status of members who regularly participate in collective activities. 
\citet{Breiger_1974} also demonstrated that the representation of a two-mode network can act as a proxy to create two different networks, where a set of actors can be connected due to a shared affiliation just like groups to which the actors belong are connected via overlapping memberships. By projecting the matrix, a rectangular matrix resulted in two square matrices. New extensions of the concept of \textsl{``duality''} aimed to go beyond structural representations and consider cultural forms such as shared objects, symbols, or expressions of taste \citep{Mutzel2020}.

Network researchers working on the sociology of science and knowledge often associate social circles -- the sociological phenomenon behind the analytical concept of \textsl{``duality''} -- with invisible colleges when analyzing researchers. 
In this literature, researchers are grouped together because they interact, have a common interest in shared topics, and do not need to know each other to be influenced by other members \citep{Kadushin_1966, Crane_1969}. 
This type of social circle requires both a social and cognitive dimension. As \citet{Zuccala_2006} clarified, an invisible college is a set of interacting scholars that share similar research interests concerning a subject specialty, as the propositional knowledge available from an intellectual group of references (``a coalition in the mind''). A subject specialty informs the invisible college of its rules and research problems and supports the intellectual motivation for social activity.
These invisible colleges should not separate the subject specialty they come from from the social aspect of science \citep{Mullins1972} such as personal communication, interactions, social relationships, groups, mentor-mentees or contexts in which knowledge takes place, making the socio-cognitive dimension of science explicit. In the postscript of \citet{Kuhn_2012}, he mentioned that paradigms are better understood when the community structure of science is taken into account, as was investigated by researchers of the time using the social network perspective (e.g., \citealp{Hagstrom_1965},\citealp{Price_Beaver_1966}, and \citealp{Crane_1969}). And \citet[p.~437]{Merton_2000} equated the concept of the ``invisible college'' \citep{Crane_1969} with socio-cognitive networks to explain the genesis and transmission of knowledge.

Duality can leverage science's socio-cognitive dimension by explicitly explaining how individuals create social ties by considering these two dimensions together. Regarding scientific networks, there is a longstanding tradition emphasizing that researchers should consider social and cognitive ties to analyze scientific networks \citep{ Crane1972, Merton_2000, White2004, White2011, Bellotti_Espinosa}. The main reason therein is that each type of tie measures something different, and separating them can lead to distorted representations of the underlying network \citep{ Holland_Leinhardt_1974, chubin1976}. Still, researchers often analyze social or cognitive ties separate from one another. For example, for \citet{Moody_2004}, citation networks are not social networks because the social ties do not capture the informational interaction structure of the latter. \citet{Schrum_Mullins_1988} distinguished between \textsl{``interactions''} and \textsl{``interest''}. The former mechanism implies communication, information flow, or general contact (such as coauthorship and \textsl{``in-house''} citation). In contrast, the latter is represented by citing the same papers (i.e., co-occurrence of citations in bibliographies). \citet{Leydesdorff_Vaughan_2006}  argues that co-occurrences in bibliometric research represent variables attributed to texts, which is different from social networks that often refer to concrete relations (such as \textsl{``affiliations''}).

\paragraph*{Collaboration and Citation.}Citations are a manifestation of formal but asymmetric communication between two scientists: if researcher $a_1$ refers to a work authored by another researcher $a_2$, it is presumed that the cited work was helpful in $ a_1$'s research \citep{chubin1976}. For \citet{Small_1978}, citations are symbols of concepts and ideas expressed in language, as cited works embody ideas that authors discuss in their work. On the other hand, to characterize the structure of a scientific field, researchers often use coauthorship networks \citep{ Newman_2001, Moody_Light_2006} as a proxy of interpersonal relationships to identify scientists' communication as a social dimension of science. These researchers distinguished between cognitive and social ties.

Separating citation and collaboration as cognitive and social dimensions added another layer of distortion by assuming the citation is cognitive without any social component interweaving with cognitive ties. The overlapping nature of these two types of relations is referred to as a socio-cognitive network \citep{Merton_2000, White2004}, and has been recently studied as a co-evolving process that allowed disentangling whether the two networks influence each other \citep{espinosa2024, lerner2024relational_coevolution} specifically by considering a delimited context. 
Researchers relied on intercitation, defined as \textsl{``the record of who has cited whom within a fixed set of authors''} \cite[p.~275]{White2011} to explore in bounded contexts, and assuming awareness, the dual nature of socio-cognitive networks such as \textsl{``in-house''} (i.e., same institution) relationships \citep{ Chubin_Studer_1979, Schrum_Mullins_1988}. 
By delimiting the context, researchers can investigate in more detail how the social dimension of scientific networks unfolds by identifying whether citations also have a social component. The resulting socio-cognitive process is inherently part of bibliometric data, leading to many possible linked networks \citep{Batagelj_2013} manifesting the duality \citep{Breiger_1974, Mutzel2020} of scientific networks. From the same scientific work, different networks can be derived, such as coauthorships, in which two authors are linked if they produce a joint work (e.g., paper, book, presentation), or a citation network, in which authors can cite many other different works, which in both cases represents a two-mode network. Bibliographic data, as a product of science, allow for the tracing of many formal communication channels in science.

\begin{figure}[t!]
    \centering
        \includegraphics[width=0.8\linewidth, page = 2,clip, trim=0.5cm 0.5cm 9cm 0cm]{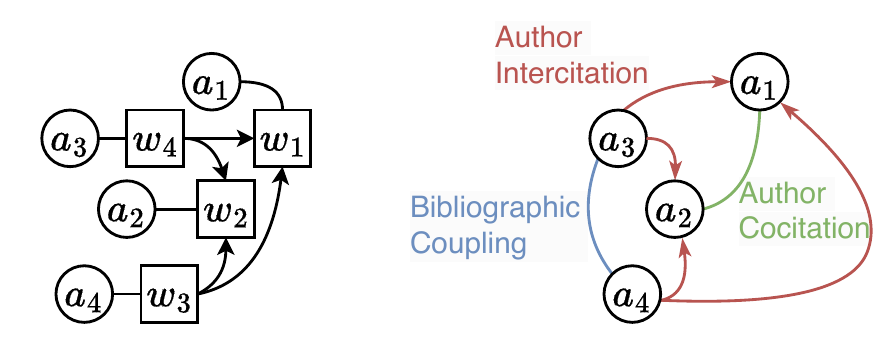}
        \caption{Illustrative examples of three types of representation. The observed citation and collaboration data are represented in the left graph. On the right, the green, red, and blue lines represent the ties resulting from author cocitation, bibliographic coupling, and author intercitation, respectively. }
        \label{fig:different_representation}
\end{figure}

\paragraph*{Types of Representations.} In the study of author-to-author citation networks, researchers often use aggregated network representations to investigate social and cognitive ties without explicitly considering the intermediate role of works. For instance, author cocitation, where \textsl{``authors whose works are generally seen to be related, and are repeatedly cited as such in later documents, tend to cluster together on the map, while authors who are rarely or never cited together are relatively far apart''} \citep[p.~164]{White_Griffith_1981}. 
While a single work is implicitly considered a building block of the representation, an author cocitation approximates how the same works cite two authors through a two-mode network. 

Another strategy that uses citation-based representations is author bibliographic coupling \citep{Zhao_Strotmann_2008}. Contrasting cocitation, two individuals are assumed to be closer if they cite the same references. In this case, the focus lies on the research front, i.e., who they cite, rather than the knowledge agreement, which relates to how other publications cite them. 

Finally, author intercitation (also referred to as author direct citation or cross-citation) is a third representation based on direct relationships between authors through citations without including a third-party work \citep{White2011}. By considering authors and works in a chain, it is apparent that for $a_1$ to cite actor $a_2$, $a_1$ must first publish work $w_2$ referencing another work $w_1$ (co-)authored by actor $a_2$ (see Fig. \ref{fig:example}). As in cocitation and bibliographic coupling, the intercitation is often treated as an aggregated matrix, assuming that the frequency of citation among $a_1$ and $a_2$ measures direct relation strength between them \citep{Wang_Qiu_Yu_2012}.

In all these representations (represented in Figure \ref{fig:different_representation}), the available timestamped interaction events are boiled down to the frequency of shared works to indicate the strength of a dyadic tie. However, treating the data as weighted ties raises new problems because works with more authors or references can overestimate their prevalence, requiring new techniques to normalize the credit given to authors (e.g., \citealp{Batagelj_2020}). At the same time, it is unlikely to explicitly address the contribution of the chain of the entire set of authors and the whole sets of works in these network representations. A representation that can explore the duality of persons and groups as pushed forward by \citet{Breiger_1974} can capture higher-order dependencies typically present in networks with these characteristics. For example, if work $w_1$ cites another work $w_2$, it may depend on the coauthorship relations between the authors of both papers. 

The aggregated representation of the ties assumes that the frequency of links between entities reflects the durability of the underlying structure of scientific networks. 
Nonetheless, analyses of the precise order and repeated interactions via authorship or citations in science have received little attention (some exceptions are \citealp{lerner2023micro} and \citealp{lerner2024relational_coevolution}).
For instance, works are events or instances in science that are scientific productions generated by an author or a team of researchers that refer to previous works by citing the references that justify the stands of the work. As \citet[p.~iii]{Garfield_1964} mentioned, the history of science depends on the sequence of events on which each discovery depends. Compared with aggregated measures, events in scientific networks allow the study of links between researchers and other entities \citep{Hummon_Doreian_1989}.

\section{Socio-cognitive Mechanisms}
\label{sec:socio-cognitive}

Researchers investigating mechanisms underlying scientific networks consider different network mechanisms \citep{Rivera_Soderstrom_Uzzi_2010, espinosa2024}. These relational mechanisms are based on dyadic similarity (e.g., homophily), relationships (e.g., Matthew effect or group structures), and proximity-based mechanisms (e.g., focuses of activity). Some mechanisms represent general patterns, while others can be dissected into concrete network representations that explicitly show entities' internal structure and relations \citep{Stadtfeld_Amati_2021}.

Socio-cognitive networks are complex structures that can be understood through the lens of mechanisms involving a mixture of different entities (e.g., authors and works) and ties (e.g., citations and collaborations). One can explore these structures in fine-grained data as micro-temporal patterns \citep{butts_relational_2023}, considering the temporal order of relational events over specific time scales. We present four general mechanisms involving socio-cognitive structures and then suggest more concrete micro-temporal mechanisms to explore these patterns.

\paragraph*{Matthew Effect of Authors.} We investigate the Matthew effect \citep{zuckerman1967nobel, merton1968matthew} by focusing on the authors of the papers as one of the primary explanations for why researchers receive more recognition over time. Originally, Harriet Zuckerman and Robert K. Merton were interested in how the allocation of credit in cases of collaboration affects the flow of ideas through the communication network of science. 
The Matthew effect highlighted the bias of allocating more recognition to renowned researchers while reducing the visibility of contributions of less well-known authors \citep{zuckerman1967nobel, merton1968matthew, Cole_Cole_1973, Merton_1988}. One instance of this mechanism is the accumulation of work citations \citep{Price_1965} and collaborators \citep{Newman_2001, barabasi2002}. 
By analyzing collaboration and citations separately, the pattern becomes a self-fulfilled prophecy in which collaborations lead to more collaborators. 
Citations of a particular work lead to more citations of that work, thus the authors of that work accumulate more recognition. 
Nonetheless, it is unclear how authors, collaboration, and citations imply cumulative processes when they are analyzed together. 
Authors can gain more recognition by receiving more citations; this accumulation might occur because a single work becomes highly visible, the author's portfolio of documents accumulates more citations, or both. 
These accumulation processes can also result from a group of researchers reinforcing the recognition or a consensus among the broader scientific community. In other words, is it the work or the author that gets repeatedly cited? 
Further, do they get cited repeatedly by the same or different authors? Due to the dual accumulation process by individual papers or the author's entire portfolio, an assortativity degree process underlies the mechanism. This is reasonable in scientific networks, as it represents the reinforcement of active actors according to the Matthew effect and their more visible positions within groups \citep{brieger1976career, mullins1977group}.

\begin{itemize}
    \item[]\textbf{Hypothesis 1 (H1):} \textsl{Actors tend to send more citations to other actors that have received more citations before.} 
\end{itemize}

\paragraph*{Intercitation.} Intercitation or author-to-author citation occurs when members of a contextually bounded group cite each other. \citet{White2004} investigate whether intercitation varies according to acquaintanceship and communication between members of these groups, intellectual affinities that are paramount regardless of the social dimension, or a combination of both.
Intercitation allows focusing on asymmetric relationships since one author citing another author does not imply that the latter cites the former. 
For intercitations, we can distinguish if an author cites another author because they are collaborators or if, because they cite each other, they will collaborate.
We further dissect this mechanism by considering the hyperevent taking into consideration author $a_1$ publishing a work $w_1$ that cites another work $ w_2 $ to a different author $ a_2 $. 

\begin{itemize}
    \item[]\textbf{Hypothesis 2a (H2a):} \textsl{Authors tend to cite other authors they had cited before.}
    \item[]\textbf{Hypothesis 2b (H2b):} \textsl{Authors tend to collaborate with other authors they had cited before.} 
    \item[]\textbf{Hypothesis 2c (H2c):} \textsl{Authors  tend to cite coauthors' works.}
\end{itemize}

\paragraph*{Author Cocitation.} \citet{White2011} challenged that intercitation was one of the main effects that explain citation because he believes that the \textsl{``true glue''} binding scientists and scholars together is what members can competently write about rather than whom they know. To him, social and affective ties are secondary to intellectual relevance. To achieve this conclusion, he explored the author cocitation mechanism as a measure that controls for the propensity of any work by an author that also appears in any work of another author to appear in the references of a later work \citep{White_Griffith_1981}. \citet{White2004} write that \textsl{``[b]ecause scholars are cited together for many reasons, cocitation data can be noisy, but in the aggregate, they are a robust measure of how citers view the intellectual linkages in a research domain''} (p. 115). This approach considers the global community, and as long as two researchers are cocited from anyone else, they would appear together in a network representation (as a symmetric weighted tie). \citet{White2004} considered that cocitation and intercitation could be conflated in bounded settings because if author $a_1$ cites himself and another author $a_2$ in the same network, it will increment the author $a_1$ to author $a_2$ intercitation and the author $a_1$ to author $a_2$ cocitation. Nonetheless, this conflation appears when self-citations are considered. 

\begin{itemize}
    \item[]\textbf{Hypothesis 3 (H3):} \textsl{Authors that are frequently cocited will tend to be cocited together more often in future publications.}
\end{itemize}

\paragraph*{Group Effect.} The last mechanism we are interested in is related to the relevance of groups. \citet{Mullins1972, Mullins1973} and \citet{Griffith_Mullins_1972} proposed that \textsl{``coherent groups''} -- which are considered small and intensely interacting research groups -- are the primary drivers of scientific change and seed larger invisible colleges that develop around them. Dyadic and triadic structures of multiple ties represent these groups, allowing the emergence of more complex morphological structures \citep{Mullins_Mullins_1973}. Recently, researchers have used similar structures by considering one layer in collaboration networks \citep{Kronegger_Ferligoj_Doreian_2011, Ferligoj_Kronegger_Mali_Snijders_Doreian_2015, Stark_Rambaran_McFarland_2020, Wittek_Bartenhagen_Berthold_2023}.

However, \citet{espinosa2024} proposed a cross-network closure that involves citation and collaboration network mechanisms. They operationalized it as \textsl{\textsl{``the tendency to cite an author if two researchers share a joint coauthor and the tendency of two actors collaborating to be cited by the same authors''}}(pp. 98-99). A further possible distinction of cross-network closure is between influence and selection processes. The ``coherent group'' can influence other members by encouraging the authors to adopt the past citations of their past coauthors or to cite the papers of their coauthors. They might also expand the group by being cocited by third authors. In both cases, the group expands by indirectly agreeing with the accumulated knowledge shared by a paradigmatic group or by directly being cocited with those with whom they agree.

\begin{itemize}
    \item[]\textbf{Hypothesis 4a (H4a):} \textsl{Coauthors  tend to cite similar references in their publications.} 
    \item[]\textbf{Hypothesis 4b (H4b):} \textsl{Coauthors tend to be cocited in subsequent publications.} 
\end{itemize}

\section{Data}
\label{sec:data}

\begin{table}[!t]
\begin{center}
\begin{tabular}{lccc}
\multicolumn{4}{l}{}                               \\ \hline
Year                                                  & 2013      & 2014      & 2015      \\ \hline
Number of authors                                           & 87       & 87       & 87       \\
Number of citing papers                                           & 322       & 345       & 367       \\
Average number of coauthors
& 2.39      & 2.18      & 2.25      \\
Average number of references
& 59.27      & 56.74      & 56.33      \\ \hline
\end{tabular}
\caption{Descriptive statistics on the number of authors and works per year in the Chileans Citation Network}
\label{table:descriptives_papers}
\end{center}
\end{table}

For the exploration of the socio-cognitive mechanisms, we make use of data involving Chilean astronomers (for details, see \citealp{Espinosa-Rada2021} and \citealp{espinosa2024}). The studied time frame corresponds to the local group formation period of astronomers and astrophysics a few years after the Atacama Large Millimeter/submillimeter Array became fully operative in 2013. The bibliographic data were initially gathered from the Web of Science, and additional information, such as the researcher's nationality, was collected from the academics' CVs. We use the Web of Science ID as a proxy of the time order,  representing the date the paper was indexed into the database. 

\paragraph*{Network Boundary.} We restrict the data to researchers affiliated with organizations settled in Chile and, thereby, have access to all the telescopes in the country. For the analysis, we consider the authors and publications in which at least one researcher is affiliated with an organization settled in Chile participating in $2013-2015$. All authors that are not settled in Chille are excluded from the analysis. 
Note, however, that foreigners can be settled in Chile. 
The cited papers are the works of this cohort published between $1947$ and $2015$.

\begin{figure}[t!]
    \centering
    \begin{subfigure}[b]{0.45\textwidth}
        \includegraphics[width=\linewidth, page = 1]{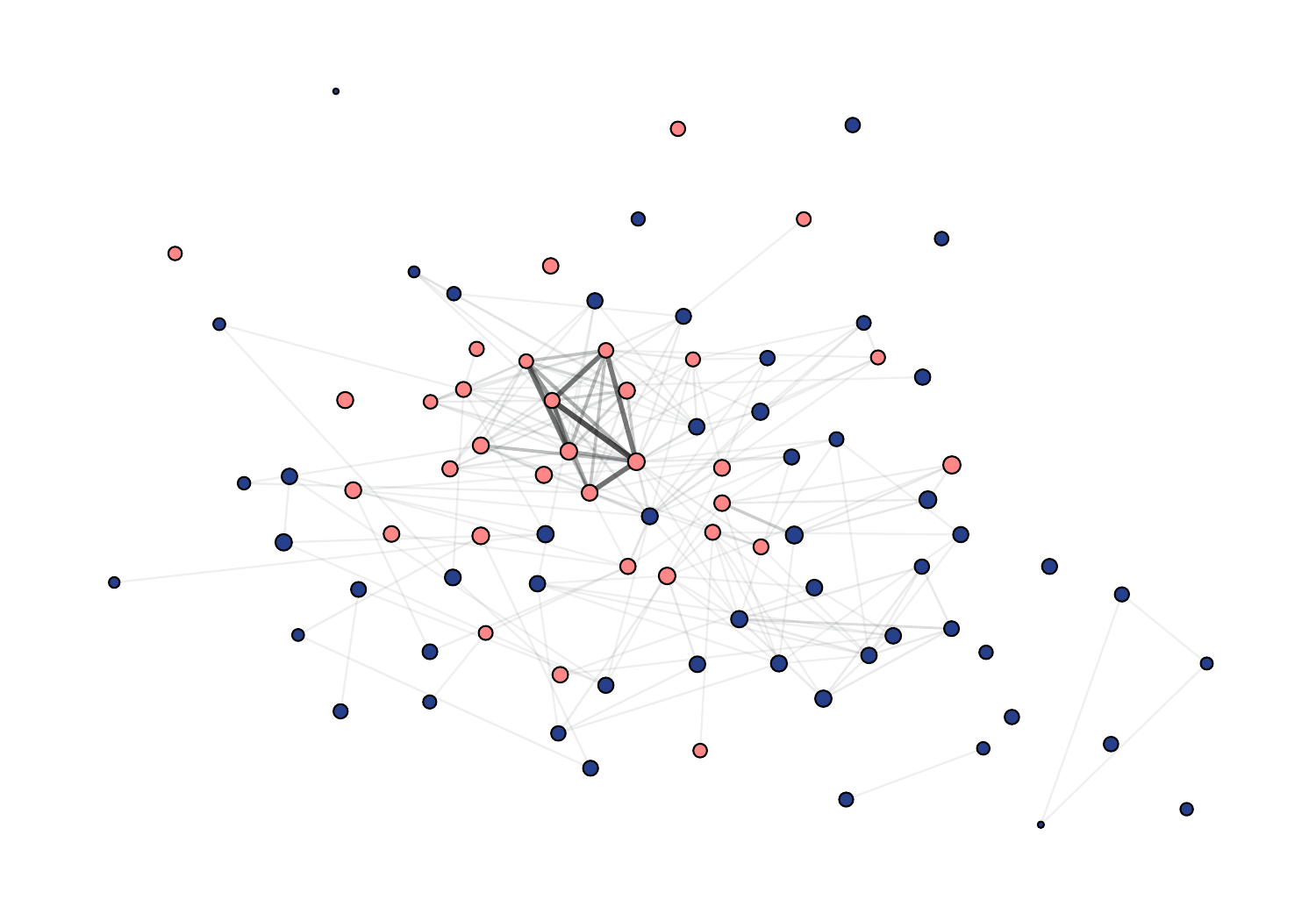} 
        \caption{Coauthorship}
        \label{fig:coauthor_network}
    \end{subfigure}
    \begin{subfigure}[b]{0.45\textwidth}
        \includegraphics[width=\linewidth, page = 1]{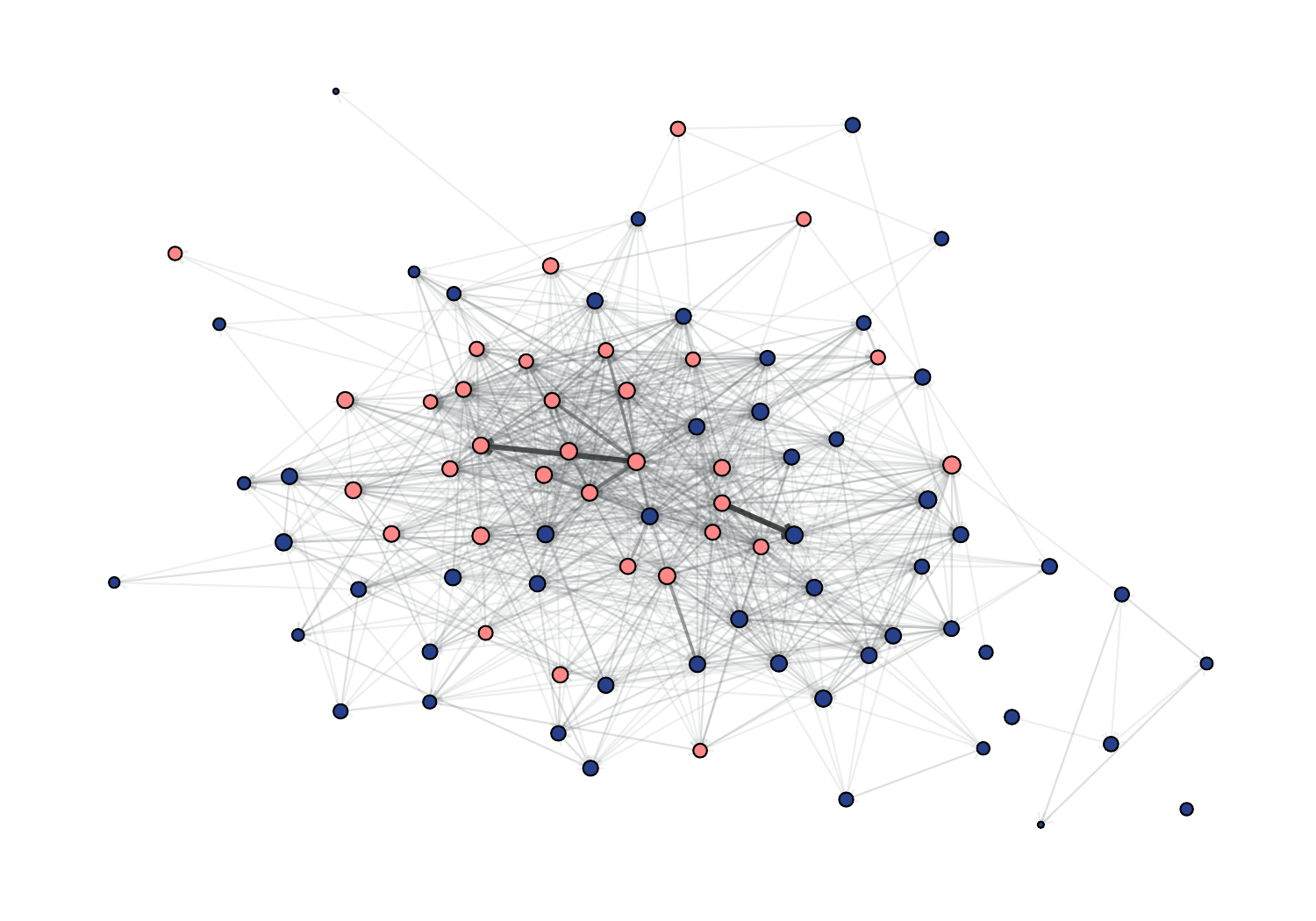}
        \caption{Citation}
        \label{fig:citation_network}
    \end{subfigure}
    \caption{Intercitation network of Chilean astronomers (2013-2015). The size of the nodes is the log-transformed number of accumulated citations, the edges are the weighted ties (number of works), and the colors represent if the nodes are Foreigners (pink) or Chileans (blue).}
    \label{fig:networks}
\end{figure}

\paragraph*{Descriptive statistics.}
 For each year, additional information on the number of authors and citing papers is given in Table \ref{table:descriptives_papers}.
The aggregated network data can be considered dual because it involves different types of nodes (works and authors) and its multiplex nature (connections based on coauthorship and citation). The two networks that distinguish between coauthor and citation relations are visualized in Fig.~\ref{fig:coauthor_network} and  Fig.~\ref{fig:citation_network}, respectively. 
While this visualization can be helpful for the exploration of the network, recent trends in network modeling for scientific networks aim to go beyond projections to analyze the two-mode structure of the network (for examples and further discussion, see \citealp{Espinosa_Ortiz_2022}, \citealp{fritz_modelling_2023}, and  \citealp{Gallagher_2023}).

\section{Methods}
\label{sec:methods}

\subsection{Group-Oriented Hyperevent Model}
\label{sec:model}

Denote the set of all $N$ astronomers by $\mathscr{A} = \{1, \ldots, N\}$ and the set of documents published up until but not including time point $t \in \mathscr{T}$ by $\mathscr{W}(t)$, with $\mathscr{T}$ being the set including all possible timestamps between 2013 and 2015 for which the bibliographic data is available. 
To shorten the notation, we assume that the static information on whether a particular astronomer in the data is Chilean is also contained in $\mathscr{W}(t)$. 
In this context, the observed coauthorship and citation data are group-to-set relational events, i.e., where the sender is a group of actors and the receiver a set of works. A publication $w = (A(w), C(w), t(w))$ encompasses an author set $A(w) \subseteq \mathscr{A}$,  citation set $C(w) \subseteq \mathscr{W}(t(w))$, and timestamp $t(w) \in \mathscr{T}$.
As described in Section \ref{sec:data}, only the order, not the exact timing of each work, is available in the publication records. 
Therefore, we assume that each work $w$ corresponds to an arbitrary timestamp $t(w)$ such that the order of the times corresponds to the observed order.
Further, $A(w)$ represents a group of authors, while $C(w)$ represents a set of works. 
Therefore, we term this type of data group-to-set relational events. 
The differentiation between groups of authors and sets of works also highlights that one can comprehend these sets as a two-mode structure of authors and works \citep{Breiger_1974} or a three-mode structure \citep{Fararo_Doreian_1984} if we include time, as they also involve authors, citing, and cited work.
The set of citable works at time point $t$ is then defined by $\mathscr{W}(t) = \{w;\, t(w) < t\}$, and the number of authors and references in $A(w)$ and $C(w)$ is denoted by $|A(w)|$ and $|C(w)|$, respectively. 
Finally, the set of sets of all possible citation lists of length $k\in \{1, 2, \ldots \}$ with information available up until but not including time point $t \in \mathscr{T}$ is given by $\mathscr{H}(t, k)$.

\paragraph*{REM for Higher-order Events.}
\citet{Lerner2021} extended the relational event model (REM) to events involving multiple actors.
The available data from Section \ref{sec:data} can, however, best be comprehended as events between groups of authors and sets of cited works, a data type for which we use the name \textsl{group-to-set events}. 
\citet{lerner2024relational_coevolution} proposed an extension of undirected hyperevents \citep{Lerner2021} and directed one-to-many hyperevents \citep{lerner2023relational_polyadic} in a one-mode network to group-to-set events in a two-mode network, which we next amend to our setting. 
 Instead of specifying the intensity of a dyadic interaction, such as author $a_1$ to cite some work by author $a_2$ at time $t$, we state a joint intensity to write a work with coauthors $A$ and citations to works $C$ at time $t$. 
 This intensity characterizes a multivariate counting process that counts how often each possible work (encompassing any number of coauthors and cited works) was written until arbitrary time point $t$. A similar model -- analyzing the interrelation between collaboration networks and references to previous work in cultural production -- has been applied by \citet{burgdorf2024communities}. We discuss alternative models in the Appendix \ref{sec:related}. 

\paragraph*{Group-Oriented Formulation.}
For the empirical setting of Chilean astronomers, we argue that coauthors affect which works are cited, but, reversely, the cited work does not affect the coauthors \citep{espinosa2024}. Therefore, we assume that the set of authors (\textsl{``sender''}) is first determined for publication, then the references (\textsl{``receiver''}) are decided upon conditional on the set of authors. 
Our model, thereby, parallels other network actor-oriented models such as the stochastic actor-oriented model \citep{Snijders_2001} or 
the dynamic network actor model \citep{stadtfeld_dynamic_2017,Stadtfeld_Block_2017}.

Mathematically, this implies a conditional independence assumption between the set of authors and its reference list, yielding the \textsl{Group-Oriented Relational Hyperevent Model}. This model comprises an author model determining the coauthors and a citation model governing the citations conditional on the set of authors and the size of the citation list. Both models take the general form of a relational event model for hyperevents involving multiple actors as proposed by \citet{Lerner2021}. Since the citation model governs the decision on a particular set of citations conditional on the set of authors and the size of the citation list, we state it as a multinomial choice model proposed by \citet{mcfadden1973}. For both models, the available information at timepoint $t\in \mathscr{T}$ includes the entire coauthorship and citation past, denoted by $\mathscr{W}(t)$.    
As a result, the intensity to observe a publication of the set of authors $A \subseteq \mathscr{A}$ with
references to $C \subseteq \mathscr{W}(t)$ at time point $t$ is given by:
\be
\label{eq:joint}
 \lambda_{A,C}(t \mid \theta, \gamma, |C|) &=& \underbrace{\lambda_{A}(t \mid \theta)}_{\text{Author Model}} \, \underbrace{p_{\,C \mid \, A}(t \mid \gamma, A, |C|)}_{\text{Citation Model}}
\ee
with 
\be
\label{eq:author}
\lambda_{A}(t \mid \theta) &=& \lambda_{0,A}(|A|,t) \exp\left(\theta^\top s(\mathscr{W}(t), A)\right)
\ee
and 
\be
\label{eq:citation}
p_{\, C}(t \mid \gamma, A, |C|) &=& \dfrac{\exp\left(\gamma^\top h(\mathscr{W}(t),C, A)\right)}{\dsum_{W \in \mathscr{H}(t, |C|)} \exp\left(\gamma^\top h(\mathscr{W}(t),W, A)\right)},
\ee
defining the author and citation model, respectively,
where 
\begin{itemize}
    \item $\theta \in \mathbb{R}^P$ and $\gamma\in \mathbb{R}^Q$ are parameter vectors estimated from data that govern the author and citation model, respectively (see Appendix \ref{sec:estimation}, for additional information on how they are estimated);
    \item $\lambda_{0}^A(|A|,t)$ is a nonparametric baseline intensity depending on the size of the author set and time $t$; 
    \item $s(\mathscr{W}(t), A)\in \mathbb{R}^P$ and $h(\mathscr{W}(t), C, A)\in \mathbb{R}^Q$ are vectors of statistics characterizing for the author and citation model, separately. These statistics detail how the intensity of observing author set $A$ and citation set $C$ given author set $A$ at time $t \in \mathscr{T}$ are affected by the bibliographic data of the past, denoted by $\mathscr{W}(t)$. This means that observing a specific group of authors or cited works depends on the authors of previous works and who they cited.
\end{itemize}

\subsection{Specification}
\label{sec:specification}
To adapt this general framework to the theory at hand, we need to specify the vectors of statistics $s(\mathscr{W}(t), A)$ and $h(\mathscr{W}(t), C, A)$ to act as proxies for the hypotheses developed in Section \ref{sec:theory} and control for other effects representing alternative explanations (or ``control effects'') for citations or coauthorship relations \citep{lerner2024relational_coevolution}. The mathematical formulation of and further details on all effects employed in our model specification are provided in Appendix~\ref{sec:sufficient_statistics}.  
Most of the hypotheses are based on the citation side; however, we also consider the author model to investigate intercitation.

\begin{figure}[t!]
    \centering
    \begin{subfigure}[b]{0.3\textwidth}
        \includegraphics[width=\linewidth, page = 1, clip, trim=0.5cm 1.5cm 0.5cm 0cm]{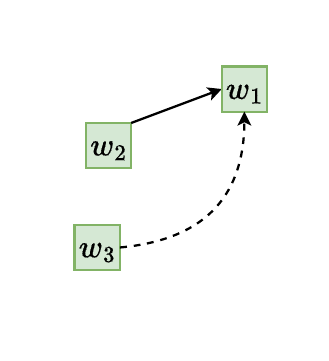} 
        \caption{\hangindent=1.9em Citation Popularity of Work}
        \label{fig:citation_pop}
    \end{subfigure}
    \begin{subfigure}[b]{0.3\textwidth}
        \includegraphics[width=\linewidth, page = 4, clip, trim=0.5cm 1.5cm 0.5cm 0cm]{Figures/all_Stats_no_caption.pdf}
        \caption{\hangindent=1.9em  Citation Repetition \newline ~ }
        \label{fig:citation_rep}
    \end{subfigure}
    \begin{subfigure}[b]{0.3\textwidth}
        \includegraphics[width=\linewidth, page = 12, clip, trim=0.5cm 1.5cm 0.5cm 0cm]{Figures/all_Stats_no_caption.pdf}
        \caption{\hangindent=1.9em Cite much Cited \newline  Authors}
        \label{fig:cite_much_cited_authors}
    \end{subfigure}   
    \caption{Statistics included for H1.}
    \label{fig:statistcs1}
\end{figure}

For H1, the accumulation of citations of works or authors may be the result of several simultaneous processes or mechanisms, which we capture by three effects illustrated in Figure \ref{fig:statistcs1}.
We consider the first two effects as control variables for basic patterns in citation events.
First, the \textsl{``Citation Popularity of Work''} (Fig. ~\ref{fig:citation_pop}) effect accounts for the effect of a paper's popularity as a process of preferential attachment. 
This effect evaluates if more frequently cited publications are more likely to receive additional citations from the academic community. 
Second, \textsl{``Citation Repetition''} (Fig. ~\ref{fig:citation_rep}) approximates the Matthew effect as a ritual process, where the same researchers repeatedly cite the same papers, typically within specific scientific specialties, promoting the cognitive group.  
Finally, we use the \textsl{``Cite much Cited Authors''} (Fig. ~\ref{fig:cite_much_cited_authors}) effect to capture the socio-cognitive structures we are interested in. 
Authors and works are interrelated, since researchers cite works of authors that have published highly cited other publications before. 


\begin{figure}[t!]
    \centering
    \begin{subfigure}[b]{0.3\textwidth}
        \includegraphics[width=\linewidth, page = 10, clip, trim=0.5cm 1.4cm 0.5cm 0cm]{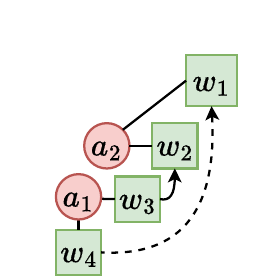} 
        \caption{\hangindent=1.9em Author cites \newline Author Repetition}
        \label{fig:author_cites_author_rep}
    \end{subfigure}
    \begin{subfigure}[b]{0.3\textwidth}
        \includegraphics[width=\linewidth, page = 11, clip, trim=0.5cm 1.5cm 0.5cm 0cm]{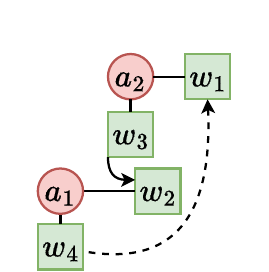} 
        \caption{\hangindent=1.9em Author cites \newline Author Reciprocity}
        \label{fig:author_cites_author_rec}
    \end{subfigure}   \\
    \begin{subfigure}[b]{0.3\textwidth}
        \includegraphics[width=\linewidth, page = 20, clip, trim=0.5cm 1.5cm 0.5cm 0cm]{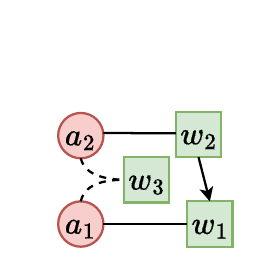}
        \caption{\hangindent=1.9emCollaboration with \newline Citing Author}
        \label{fig:collaborate_with_citing_author}
    \end{subfigure}  
    \begin{subfigure}[b]{0.3\textwidth}
        \includegraphics[width=\linewidth, page = 9, clip, trim=0.5cm 1.5cm 0.5cm 0cm]{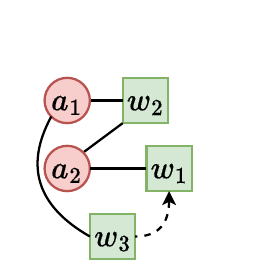}
        \caption{\hangindent=1.9em Cite Coauthor Works\newline ~ }
            \label{fig:cite_coauthor_papers}
    \end{subfigure}  

    \caption{Statistics included for H2.}
    \label{fig:statistcs2}
\end{figure}

The following effects approximate the intercitation mechanism. The first effect, \textsl{``Author cites Author Repetition''} (Fig. ~\ref{fig:author_cites_author_rep}), tests H2a and controls for researchers' inclination to follow the work of prominent figures, often leaders in their specialties, by promoting their research agenda \citep{Mullins_Mullins_1973}. The second effect, \textsl{``Author cites Author Reciprocation''} (Fig. ~\ref{fig:author_cites_author_rec}), also complements the first effect by incorporating reciprocity as a controlling effect, to evaluate if there is mutual admiration or recognition that may occur among researchers interested in similar topics.
A third effect for H2b considers the author's model by investigating whether scientists are more likely to coauthor papers with those who cited their previous work (\textsl{``Collaborate with Citing Author''} in Fig. ~\ref{fig:collaborate_with_citing_author}). 
Finally, for H2c, we employ the \textsl{``Cite Coauthor Works''} effect (Fig. ~\ref{fig:cite_coauthor_papers}), which can be comprehended as an effect reversing the temporal order of \textsl{``Collaborate with Citing Author''} in that scientists are first coauthors and then cite each other. 

\begin{figure}[t!]
    \centering
    \begin{subfigure}[b]{0.3\textwidth}
        \includegraphics[width=\linewidth, page = 14, clip, trim=0.5cm 1.5cm 0.5cm 0cm]{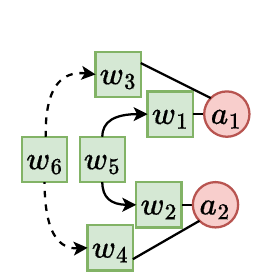}
        \caption{\hangindent=1.9emAuthor Cocitation\newline ~}
        \label{fig:author_co_citation}
    \end{subfigure}  
    \begin{subfigure}[b]{0.3\textwidth}
        \includegraphics[width=\linewidth, page = 2, clip, trim=0.5cm 1.5cm 0.5cm 0cm]{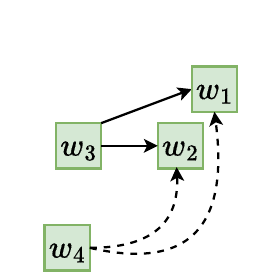}
        \caption{\hangindent=1.9em Cocitation \newline Popularity: Pair}
        \label{fig:co_citation_pair}
    \end{subfigure}  
    \begin{subfigure}[b]{0.3\textwidth}
        \includegraphics[width=\linewidth, page = 3, clip, trim=0.5cm 1.5cm 0.5cm 0cm]{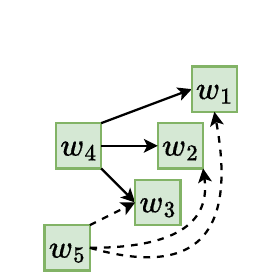}
        \caption{\hangindent=1.9em Cocitation \newline  Popularity: Triplet}
        \label{fig:co_citation_tri}
    \end{subfigure}  
    \caption{Statistics included for H3.}
    \label{fig:statistcs3}
\end{figure}

As in H1, we use three effects to investigate H3. The first effect (Fig. ~\ref{fig:author_co_citation}) represents the hypothesis related to author cocitation (H3), and the other two (Fig. ~\ref{fig:co_citation_pair} and Fig. ~\ref{fig:co_citation_tri}) control for simpler explanations for cocitation, which are lower-order terms. The effect of \textsl{``Author Cocitation''} (Fig. ~\ref{fig:author_co_citation}) aims to identify the recurrence of authors perceived as working on similar topics by later publications. However, in the case of \textsl{``Cocitation Popularity (Pair)''} (Fig. ~\ref{fig:co_citation_pair}) and \textsl{``Cocitation Popularity (Triple)''} (Fig. ~\ref{fig:co_citation_tri}), instead of using author cocitation \citep{White_Griffith_1981}, we control for standard cocitation at the level of works \citep{Small_1973}. The main difference is that in the former case, we emphasize the duality of authors and works, while in the latter, the focus is on the cognitive dimension without considering which authors are behind the publication.

\begin{figure}[t!]
    \centering
     \begin{subfigure}[b]{0.3\textwidth}
        \includegraphics[width=\linewidth, page = 8, clip, trim=0.5cm 1.5cm 0.5cm 0cm]{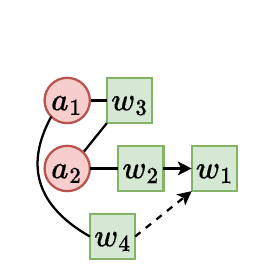}
        \caption{\hangindent=1.9em Adopt Coauthor \newline Citation}
        \label{fig:adopt_coauthor_citations}
    \end{subfigure}  
    \begin{subfigure}[b]{0.3\textwidth}
        \includegraphics[width=\linewidth, page = 13, clip, trim=0.5cm 1.5cm 0.5cm 0cm]{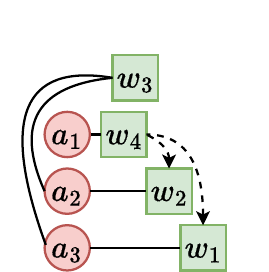}
        \caption{\hangindent=1.9em Cocite Coauthor Pair\newline ~}
        \label{fig:cite_coauthor_pair2}
    \end{subfigure} 
    \caption{Statistics included for H4.}
    \label{fig:statistcs4}
\end{figure}

We use two variants specified in H4a and H4b to investigate the described group mechanisms. We test hypothesis H4a using \textsl{``Adopt Coauthor Citation''} (Fig. ~\ref{fig:adopt_coauthor_citations}), which is a \textsl{``social influence''} effect in which scientists cite some of the works cited by their former coauthors. To test H4b, we consider the statistic \textsl{``Cocite Coauthor Pair''}(Fig. ~\ref{fig:cite_coauthor_pair2}), which represents the pattern of cociting papers of a pair of coauthors.

\begin{table}[!tbp]
\begin{center}
\caption{Operationalization of the main Hypotheses.\label{tbl:opera}}
\resizebox{\textwidth}{!}{
\begin{tabular}{llll}
Hypothesis                         & Sub-hypothesis & Effects                            & Model    \\ \hline
H1: Matthew effect of Authors      & H1 (Fig. \ref{fig:cite_much_cited_authors})            & Cite much Cited Authors            & Citation \\\hline
\multirow{4}{*}{H2: Intercitation} & H2a (Fig. \ref{fig:author_cites_author_rep})           & Author cites Author Repetition     & Citation  \\
                                   
                                   & H2b (Fig. \ref{fig:collaborate_with_citing_author})           & Collaborate with Citing Author     & Author   \\
                                   & H2c (Fig. \ref{fig:cite_coauthor_papers})            & Cite Work of Coauthor              & Citation \\\hline
H3: Author Cocitation              & H3  (Fig. \ref{fig:author_co_citation})            & Author cocitation                  & Citation \\ \hline
\multirow{2}{*}{H4: Group effect}  & H4a     (Fig. \ref{fig:adopt_coauthor_citations})       & Adopt Citation of Coauthor         & Citation \\
                                   & H4b        (Fig. \ref{fig:cite_coauthor_pair2})    & Cocite Coauthor Pairs              & Citation \\

\end{tabular}
}
\end{center}
\end{table}

Table~\ref{tbl:opera} summarizes the main hypotheses and effects and to which model each effect corresponds.
In the author model, we include additional control effects such as the ratio of Chileans in author teams compared to foreigners (``Ratio Chileans'') and the heterogeneity of the team of coauthors concerning Chilean nationality (``Heterogeneity Chilean''). We also consider some degree-based effects such as ``Citation Popularity of Authors'', ``Publication Activity'', and ``Coauthor Repetition'' (for pairs, triples, and quartets of authors). Additionally, we explore other transitivity-based effects for social (``Closure by Coauthor'') and cognitive (``Closure by Citing Same Work'') structures. In the citation model, we account for further effects including ``Outdegree Popularity'', which controls for the tendency to cite papers with long reference lists, ``Cite Work and its Citations,'' as an appropriation of knowledge from the baseline publication, and ``Self-Citation'', as the effect of authors citing their own past work.

\section{Results}
\label{sec:results}
Before presenting the model results, note that the coefficients $\theta$ and $\gamma$ can be interpreted similarly to those in proportional hazard models. 
For $p \in {1, \ldots, P}$, consider two distinct author sets $A \subseteq \mathscr{A}$ and $A^\star \subseteq \mathscr{A} $. 
Let $\theta_p$ and $s_p(\mathscr{W}(t), A)$ denote the  $p$th coefficient and statistic in \eqref{eq:author}. 
If $s_k(\mathscr{W}(t), A) = s_k(\mathscr{W}(t), A^\star)$ for all  $k \neq p$  and $s_p(\mathscr{W}(t), A) = s_p(\mathscr{W}(t), A^\star) + 1$, then $\theta_p > 0$ indicates that  $A^\star$  is more likely than  $A$ . 
The coefficient $\gamma$ can be interpreted in a similar manner. Further details are provided in Appendix \ref{sec:interpretation}.

\subsection{Author model}
The author model explains the set of authors $A(w)$ of the published work $w$, irrespective of the citations $C(w)$. 
Findings are given in the upper part of Table~\ref{tbl:res_citation}.
We also assess for each included covariate the differences in the Akaike Information Criterion (AIC -- negative values indicate improvement) over the null model and in the full model. 
To measure contributions over the null model, we compare the AIC of the model containing just the focal effect with the AIC of the model containing only a baseline intensity. 
To measure contributions in the full model, we compare the AIC of the model containing all effects (that is, all effects of the author model or all effects of the citation model described in Section~\ref{sec:sufficient_statistics}) with the AIC of the model containing all effects but the focal one. 
The upper part of Table~\ref{tab:aic_contributions_order_null} presents the differences in AIC values in the author model, ordered by their contributions over the null model.

\begin{table}[!t]
\caption{Results: The first column shows the estimated coefficients, 
      the second column the standard errors, and the third column the p-values.
       Note that H1 stands for Hypothesis 1, H2 for  Hypothesis 2 and so on.
       \label{tbl:res_citation}} 
\begin{center}
\begin{tabular}{lrrr}
\hline
\multicolumn{1}{l}{}&\multicolumn{1}{c}{Estimate}&\multicolumn{1}{c}{Std. Error}&\multicolumn{1}{c}{p-Value}\tabularnewline
\hline
{\  Author Model}&&&\tabularnewline 
~~Ratio Chileans&$-0.166$&$0.074$&0.026\tabularnewline 
~~Heterogeneity Chilean&$ 0.791$&$0.168$&\textless 0.001\tabularnewline 
~~Citation Popularity of Author&$-0.320$&$0.074$&\textless 0.001\tabularnewline 
~~Publication Activity&$0.506$&$0.049$&\textless 0.001\tabularnewline 
~~Coauthor-pair Repetition&$ 1.157$&$0.043$&\textless 0.001\tabularnewline 
~~Coauthor-triple Repetition&$ 0.224$&$0.011$&\textless 0.001\tabularnewline 
~~Coauthor-quartet Repetition&$ 0.070$&$0.009$&\textless 0.001\tabularnewline 
~~\textbf{H2b:} Collaborate with Citing Author&$ 0.114$&$0.046$&0.013\tabularnewline 
~~Closure by Coauthor&$-0.861$&$0.082$&\textless 0.001\tabularnewline 
~~Closure by Citing same Work&$-0.115$&$0.029$&\textless 0.001\tabularnewline 
\hline
{\  Citation Model}&&&\tabularnewline 
~~Citation Popularity of Work&$-0.153$&$0.140$&0.274\tabularnewline 
~~Cocitation Popularity: Pair&$ 0.265$&$0.029$&\textless 0.001\tabularnewline 
~~Cocitation Popularity: Triple&$ 0.073$&$0.005$&\textless 0.001\tabularnewline 
~~Citation Repetition&$ 0.151$&$0.047$& 0.001\tabularnewline 
~~Outdegree Popularity&$-0.620$&$0.103$&\textless 0.001\tabularnewline 
~~Cite Work and its Citations&$ 0.297$&$0.021$&\textless 0.001\tabularnewline 
~~Self Citation&$ 1.531$&$0.115$&\textless 0.001\tabularnewline 
~~\textbf{H1:} Cite much Cited Authors&$-1.388$&$0.210$&\textless 0.001\tabularnewline 
~~\textbf{H2a:} Author cites Author Repetition&$ 0.730$&$0.209$&\textless 0.001\tabularnewline 
~~Author cites Author Reciprocation&$-0.112$&$0.206$&0.585\tabularnewline 
~~\textbf{H2c:} Cite Work of Coauthor&$-0.384$&$0.138$&0.006\tabularnewline 
~~\textbf{H3:} Author Cocitation&$-0.165$&$0.109$&0.130\tabularnewline 
~~\textbf{H4a:} Adopt Citation of Coauthor&$-0.063$&$0.079$&0.422\tabularnewline 
~~\textbf{H4b:} Cocite Coauthor Pairs&$ 0.480$&$0.110$&\textless 0.001\tabularnewline 
\hline
\end{tabular}\end{center}
\end{table}

\begin{table}[!t]
\caption{AIC Contributions: Over null models and in full models. More negative values point to stronger contributions. Effects are ordered by the contributions over the null models. The percentages in brackets give the share of the AIC differences between the respective full models and null models. In the Author Model, the AIC of the full model minus the AIC of the null model is $-$2629.438, corresponding to 100\%. In the Citation Model, the AIC of the full model minus the AIC of the null model is $-$13914.82, corresponding to 100\%.}

\begin{tabular}{lrr}
\hline
\multicolumn{1}{l}{Model Term}&\multicolumn{1}{c}{over null model}&\multicolumn{1}{c}{in full model}
\\ \hline
\multicolumn{3}{l}{Author Model} \\ 
~~Coauthor-pair Repetition & $-$1507.248 (57.3\%)& $-$593.090  (22.6\%)\\ 
~~Closure by Citing same Work & $-$997.218  (37.9\%)& $-$12.278  (0.5\%)\\ 
~~Coauthor-triple Repetition & $-$825.247  (31.4\%)& $-$263.097  (10.0\%)\\ 
~~Publication Activity & $-$763.299  (29.0\%)& $-$101.118  (3.8\%)\\ 
~~\textbf{H2b:} Collaborate with Citing Author & $-$760.261  (28.9\%)& $-$3.890  (0.1\%)\\ 
~~Closure by Coauthor & $-$593.533  (22.6\%)& $-$200.996  (7.6\%)\\ 
~~Citation Popularity of Author & $-$472.499  (18.0\%)& $-$20.720  (0.8\%)\\ 
~~Coauthor-quartet Repetition & $-$371.534  (14.1\%)& $-$81.990  (3.1\%)\\ 
~~Ratio Chileans & $-$110.515  (4.2\%)& $-$3.217  (0.1\%)\\ 
~~Heterogeneity Chilean & 0.198  ($-$0.0\%)& $-$20.708  (0.8\%)\\ \hline
\multicolumn{3}{l}{Citation Model} \\ 
~~Self citation & $-$11657.594 (83.8\%)& $-$1340.308 (9.6\%)\\ 
~~Citation Repetition & $-$6920.160 (49.7\%)& $-$21.127 (0.2\%)\\ 
~~Cocitation Popularity: Triple& $-$6315.904  (45.4\%)& $-$249.832 (1.8\%)\\ 
~~Cocitation Popularity: Pair & $-$4813.124  (34.6\%)& $-$187.990 (1.4\%)\\ 
~~Cite work and its Citations & $-$4785.378  (34.4\%)& $-$360.689 (2.6\%)\\ 
~~\textbf{H2a:} Author cites Author Repetition& $-$2967.863  (21.3\%)& $-$30.325 (0.2\%)\\ 
 ~~Author cites Author Reciprocation & $-$2637.315  (19.0\%)& 1.258 ($-$0.0\%)\\ 
~~\textbf{H4a:} Adopt Citation of Coauthor & $-$1235.190  (8.9\%)& 0.847 ($-$0.0\%)\\ 
~~\textbf{H2c:} Cite work of Coauthor & $-$568.488 (4.1\%)& $-$17.708 (0.1\%)\\ 
~~Citation Popularity of Work & $-$525.869 (3.8\%)& $-$1.731 (0.0\%)\\ 
~~\textbf{H4b:} Cocite Coauthor Pairs & $-$172.407 (1.2\%)& $-$35.353 (0.3\%)\\ 
~~\textbf{H3:} Author Cocitation & $-$16.773 (0.1\%)& $-$1.999 (0.0\%)\\ 
~~\textbf{H1:} Cite much Cited Authors & 0.205 ($-$0.0\%)& $-$102.173 (0.7\%)\\ 
~~Outdegree Popularity & 0.942 ($-$0.0\%)& $-$78.691 (0.6\%)\\ \hline
\end{tabular}
\label{tab:aic_contributions_order_null}
\end{table}

\paragraph*{Hypotheses Effects.} Regarding H2b, we find a positive effect to \textsl{``Collaborate with Citing Author''}. 
This suggests that scientists tend to coauthor works with those who cited their own work in the past, as predicted by H2b. That is, there is a social selection of coauthors having cited their own work. From another point of view, there is a cross-network effect in the sense that if author $a_1$ has cited the work of author $a_2$, then it is more likely that $a_1$ and $a_2$ become coauthors in the future. The effect to collaborate with an author who has cited the own work makes a contribution of intermediate strength over the null model, compared to the contributions of the control effects. However, its contribution in the full model is much weaker, albeit it leads to a model improvement.

\paragraph*{Other Effects.} Foreigners are more likely to be included in sets of authors than Chileans (negative effect of \textsl{``Ratio Chilean''}, significant at the 10\% level), and groups of coauthors are more heterogeneous than groups of randomly sampled scientists. That is, teams of coauthors are often mixed with Chileans and foreigners (positive effect of \textsl{``Heterogeneity Chilean''}).

There is a positive effect of the number of citations that an author's works have received in the past (\textsl{``Citation Popularity''}). That is, those scientists whose works have been cited more in the past publish at a higher rate in the future, i.e., they are more likely to be included as coauthors. There is a negative effect of \textsl{``Publication Activity''} on the publication rate. Which, scientists who have published more in the past will publish at a lower rate in the future. This latter effect works towards equalizing publication counts in the population of scientists.

There is evidence for \textsl{``Coauthorship Repetition''} among groups of scientists of sizes two, three, and four. That is, those groups of the given sizes who have coauthored more in the past are more likely to coauthor work in the future.

There is a significant negative triadic closure (\textsl{``Closure by Coauthor''}) in the coauthoring network. According to \citet{lerner2022dynamic,lerner2023micro}, negative closure points to actors occupying stable broker positions (actors surrounded by structural holes, that is, actors bridging between communities). A positive closure effect would imply a tendency to close structural holes. Still, the negative closure effect found in our model suggests that authors are likely to keep structural holes open so that the ``third author'', i.\,e., the one connected to the two others, is likely to keep her broker position. Likewise, if two scientists cited the same works, they are less likely to jointly publish a work (\textsl{``Closure by Citing Same Work''}).

\subsection{Citation Model}

The citation model explains the list of citations $C(w)$ of a published work $w$, conditioning on its group of authors $A(w)$. Findings are given in the lower part of Table \ref{tbl:res_citation}. The lower part of Table~\ref{tab:aic_contributions_order_null} presents the differences in AIC values implied by the various effects, ordered by their contributions over the null model.

\paragraph*{Hypotheses Effects.}
Hypothesis H1 is represented by the effect \textsl{``Cite Much Cited Authors''}, displayed in Fig.~\ref{fig:cite_much_cited_authors}. 
In the example given in that figure, $a_2$ authored the work $w_1$, which received many citations in the past, and $a_2$ also authored work $w_2$. 
H1 predicts that there is in increase probability of $w_2$ being cited in the future. Put differently, we expect a \textsl{``spill-over''} effect of the popularity of $w_1$ onto the work $w_2$ written by the same author $a_2$. 
Contrary to these expectations, we find a negative effect of \textsl{``Cite Much Cited Authors''}. 
In the setting above, this suggests that $w_2$ gets cited at a lower rate in the future. 
Still, this finding must be interpreted alongside several other effects that control for repeatedly citing the same works or the same authors:
We find a positive tendency for \textsl{``Citation Repetition''}, displayed in Fig.~\ref{fig:citation_rep}, suggesting that if author $a_1$ has already cited work $w_1$, then the same author $a_1$ is more likely to cite $w_1$ again when publishing another paper. 
\hide{
We also find a positive effect for \textsl{``Author Cites Author Repetition''} (Fig.~\ref{fig:author_cites_author_rep}), suggesting that if $a_1$ has already cited work $w_2$ of author $a_2$, then the same author $a_1$ is more likely to cite another paper ($w_1$) of $a_2$ in the future. 
}
The second control effect, \textsl{``Citation Popularity of Work''}, displayed in Fig.~\ref{fig:citation_pop}, has a negative parameter (significant at the 10\% level). This suggests that works that received many citations in the past get cited at a lower rate in the future, controlling for all other effects. 
The contribution of the effect \textsl{``Cite Much Cited Authors''} to the AIC is of moderate strength in the full model -- compared to the contributions of other effects -- but its contribution over the null model is non-existent. 
In general, we observe that -- especially in the references model but also in the authors model -- several of the control variables make stronger contributions than any of the effects related with our hypotheses. 
This does not come as a surprise and, when interpreting results, we have to distinguish between ``\textsl{explaining the data}'' and ``\textsl{testing relevant hypotheses}''. 
Indeed, for example, the tendency of authors to cite their own work (\textsl{``Self citation''}) or to repeatedly cite the same papers (\textsl{``Citation repetition''}) are fairly obvious patterns that empirically explain a large share of the variance in the data; on the other hand there is hardly any novel or unexpected insight coming from these patterns.

Hypothesis H2a is represented by \textsl{``Author Cites Author Repetition''} (Fig.~\ref{fig:author_cites_author_rep}) and we also control for \textsl{``Author Cites Author Reciprocation``} (Fig.~\ref{fig:author_cites_author_rec}). Consistent with H2a, we find a positive effect of repetition, implying that if author $a_1$ has cited work of $a_2$, then $a_1$ is more likely to cite a (possibly different) work of $a_2$ in the future. There is no significant finding for reciprocation. The latter effect predicts that if author $a_1$ has cited work of $a_2$, then $a_2$ is more likely to cite work of $a_1$ in the future; thus, it reverses the roles of the citing and cited authors. Looking at the contributions to the AIC of these effects we find that both make an intermediate contribution over the null model and \textsl{``Author Cites Author Repetition''} makes a small but positive contribution in the full model, while \textsl{``Author Cites Author Reciprocation''} entails no improvement in the full model (consistent with its statistical insignificance). 

Hypothesis H2c is represented in Fig.~\ref{fig:cite_coauthor_papers} \textsl{``Cite Work of Coauthor''}. In the example given in this figure, $a_1$ and $a_2$ have coauthored work $w_2$ and $a_2$ has published $w_1$. Given this precondition, H2c predicts that $a_1$ is more likely to cite the work $w_1$ of her coauthor in the future. Contrary to the predictions of H2c, we find a negative effect to cite the work of former coauthors. We recall that H2b, \textsl{``Collaborate with Citing Author''}, has been (positively) tested with the author model, discussed above. The contribution to the AIC of \textsl{``Cite Work of Coauthor''} entails a small to intermediate improvement over the null model and a small but positive improvement in the full model.

Hypothesis H3 is represented by \textsl{``Author Cocitation''}, displayed in Fig.~\ref{fig:author_co_citation}. This effect predicts that if (possibly different) works of authors $a_2$ and $a_3$ jointly appear in the reference list of a past work $w_5$, then a future work $w_6$ is more likely to cocite (possibly yet other and possibly different) works of $a_2$ and $a_3$. With the effects displayed in Figs.~\ref{fig:co_citation_pair} and~\ref{fig:co_citation_tri} -- cocitation popularity of pairs and triples of works -- we control for the baseline effect to repeatedly cocite the same (pairs or triples of) works, rather than to cocite different works of the repeated pair of authors. Contrary to the predictions of H3, we find a negative effect (significant at the 10\% level) to repeatedly cocite authors. In contrast, repeated cocitation to pairs and triples of papers is significantly positive. The contributions to the AIC over the null model or in the full model of \textsl{``Author Cocitation''} are almost inexistent, while the contributions of the cocitation popularity of papers are very strong.

Hypothesis H4a is represented in Fig.~\ref{fig:adopt_coauthor_citations} \textsl{``Adopt Citations of Coauthors''} and H4b is represented in Fig.~\ref{fig:cite_coauthor_pair2} \textsl{``Cocite Coauthor Pairs''}. In the example given in these figures, \textsl{``Adopt Citations of Coauthors''} has as precondition that $a_1$ and $a_2$ are coauthors (having coauthored work $w_3$) and $a_2$ has cited $w_1$ when publishing $w_2$. Given this precondition, $a_1$ is predicted to be more likely to also cite work $w_1$ when publishing another future work $w_4$. However, we find no significant effect in adopting citations of coauthors, contrary to the predictions of H4a. The Effect \textsl{``Cocite Coauthor Pairs''} has as a precondition that $a_2$ and $a_3$ have coauthored work $w_3$ and have individually published $w_2$ and $w_1$. Given this precondition, the effect predicts that $w_1$ and $w_2$ have an increased probability to be cocited by the future work $w_4$. Consistent with the predictions of H4b we find a positive effect to cocite works individually published by former pairs of coauthors. Consistent with the significance or insignificance of their parameters \textsl{``Adopt Citations of Coauthors''} entails no improvement to the AIC in the full model while \textsl{``Cocite Coauthor Pairs''} entails a small improvement. Both effects make small to moderate contributions over the null model.

\paragraph*{Other Effects.} We find a positive tendency to \textsl{``Cite Work and its Citations''}. There are different possible explanations of this effect, as discussed in \citet{lerner2024relational_coevolution}. Among others, it may be that authors copy parts of the reference lists of works they cite; that authors search for citations to a work they cite (and subsequently cite some of the citing works); or that work $w_2$ is topically similar to the work $w_1$ it cites, increasing the probability that $w_2$ and $w_1$ get cocited in the future. Works with longer reference lists -- Effect \textsl{``Outdegree Popularity''} -- are less likely to be cited. Moreover, authors frequently cite their own works (\textsl{``Self Citation''}).

With respect to the contribution of the other effects over the null model and in the full model (Table~\ref{tab:aic_contributions_order_null}), we have already noted above that some controlling effects are more prominent in both cases. This is the case of coauthor-pair repetition, coauthor-triple repetition, and publication activity for the author model, and self citation (the most noticeable effect in the analysis), cocitation popularity triple, and cocitation popularity pair for the citation model. Regarding our main hypotheses, in both cases, some intercitations effects (for H2a: author cites author repetition and H2c: cite work of coauthor) are consistently contributing over the null model and in the full model. For the case of the Matthew effect of authors (H1), when we consider its contribution to the full model, we observe that the effect is more prominent than the other hypotheses (but relatively low in the null model) followed by the group effect to cocite coauthor pairs (H4b). The importance of these effects in the full model reveals the relevance of considering collaboration and citation events embedded in higher-order relations, that is, embedded in a hypergraph. Finally, the contribution of the author's cocitation (H3) effect in both cases is fairly modest in comparison with the other effects. 

\section{Discussion}
\label{sec:discussion}
Using bibliometric data to analyze a scientific community of astronomers in Chile, we investigated social and cognitive ties. By defining the context, we explored the social dimension—often linked to invisible colleges \citep{Crane1972, Zuccala_2006} but also to underlying knowledge dissemination \citep{carley1986, collins2002}—as a dual process involving both authors and their works \citep{Bellotti_Espinosa}. This approach follows the tradition of \textsl{``duality''} \citep{Breiger_1974, Mutzel2020}, explicitly considering science’s structural and cultural forms as socio-cognitive networks. Examining these processes and the micro-temporal mechanisms at play allowed us to identify patterns that influence authors' citation behaviors. While citation is often used to assess researchers' impact, it also plays a broader role in shaping the stratification of science.

While the literature consistently states that social stratification exists in science \citep{Price_1965, Cole_Cole_1973, Newman_2001, barabasi2002}, our results indicate that authors do not preferentially cite those who have received more citations in the past. These findings highlight the benefits of our approach in not projecting the data, as they show that although citation repetition occurs between, it is more predominant at the level of works rather than authors. Although stratification may be more apparent in larger networks, authors in this bounded community do not accumulate recognition and are cited at a lower rate over time. Similarly, works that have received many citations in the past tend to be cited less frequently in the future. Nonetheless, the accumulation of recognition through citation is linked to previously cited works. Thus, we interpret these findings as evidence of a preference for contemporaneous work among Chilean astronomers in scientific knowledge.

Regarding the preference for recent research among Chilean astronomers, many important control effects do not take the cited authors into account, as shown in Table \ref{tab:aic_contributions_order_null}. This phenomenon suggests that subject specialization, sets of beliefs and bodies of knowledge, is a key dimension underlying this community. One instance of this is citation repetition, where certain papers function as accumulative processes in which future work builds on previous research. Moreover, we obtain positive coefficients for the cocitation popularity of pairs and triples, reflecting cognitive closure processes at the level of scientific knowledge (i.e., works). Some authors, from an intercitation perspective, are still cited repetitively and collaborate with citing authors, but the coefficient for citing highly cited authors is negative. We interpret the relevance of intrinsic ideas, regardless of authorship, as an indication of a tendency toward specialization.

For socio-cognitive ties, \citet{White2011} argues that the \textsl{“true glue”} binding scientists and scholars together is what they can competently write about rather than whom they know. Recent research \citep{espinosa2024} has challenged this claim by reaffirming the relevance of the social dimension in the study of socio-cognitive networks. However, our findings differ from those of \citet{espinosa2024}. While their study found that prior collaboration leads to citation but prior citation does not lead to collaboration, we observe the opposite effect. Similarly, \citet{lerner2024relational_coevolution} found that scientists tend to coauthor with those who have cited their work and are inclined to cite their coauthors’ papers.

Our results indicate that authors are more likely to collaborate with individuals who have previously cited their work. Conversely, they tend to cite the work of former coauthors less frequently in subsequent research compared to non-collaborators. Some of these patterns may be linked to implicit local hierarchy practices. Controlling for other effects, we find that scientists tend to cite their former coauthors or highly cited authors less frequently than others. However, these results must be interpreted in light of a strong positive effect of citation repetition, which reflects a lower-order representation of the Matthew effect.

We conjecture that unobserved status-related processes—such as academic seniority or forms of reputation not captured by citation counts—may be shaping citation dynamics. The divergence of our findings from previous expectations highlights the need for further investigation.

The prominence of ideas is also observed in the case of cocitation, which is considered a key driver of why researchers are cited. Classical research \citep{Small_Griffith_1974} established that cocitation identifies relationships between works deemed important by authors within a specialty. \citet{White_Griffith_1981} extended this concept to author cocitation, showing how authors whose works are generally regarded as related tend to cluster in knowledge maps. Our results indicate that authors frequently cited together do not tend to be cited more often in future publications. A similar effect is observed when considering repeated cocitation at the level of works (pairs and triples) rather than authors. Both effects make a strong contribution to the author and citation model shown in Table \ref{tab:aic_contributions_order_null}. These findings suggest that in the local astronomical community, cocited authors tend to be less prominent than cocited works in shaping the development of scientific specialties.

We also highlight the importance of exploring the role of 'coherent groups'—groups that tend to expand over time. These groups have historically been considered crucial for explaining how scientific specialties evolve \citep{Mullins_Mullins_1973}. More recently, research on scientific networks suggests that science is becoming increasingly team-oriented \citep{Wuchty_Jones_Uzzi_2007, Jones_Wuchty_Uzzi_2008, Leahey_2016}. We build on this perspective by explicitly examining the effect of these groups on citation patterns, focusing on dualities in the interplay of works and authors. This is particularly relevant, as some researchers have argued that, in such communities, a researcher's group of colleagues is the most significant source of social influence on their work \citep{Hagstrom_1965}. Our findings indicate that collaborating authors do not tend to cite similar references in their publications. However, authors who collaborate are more likely to be cocited in subsequent publications. This effect suggests that the astronomical community is more inclined to cite dyadic teams, as our analysis did not account for triadic or higher-order structures.

From a methodological perspective, we contribute to the study of scientific networks by modeling citation dynamics without aggregating data at the author or paper level—a common practice in empirical studies and classical two-mode projections, such as those proposed by \citet{Breiger_1974}. Instead, we model raw group-to-set citation events directly. This approach enhances the capacity of group-oriented relational event models.

We demonstrate the utility of this method through an empirical case study and further extend the model by incorporating novel structural effects that capture dependencies among overlapping sets of nodes and ties. Our framework allows for a more granular analysis of citation mechanisms, enabling the investigation of dualities, such as the interplay between authors and works through (co)authorships or citation networks. The Group-Oriented Relational Hyperevent Model provides a way to model hyperevents in situations where groups are first formed and then decide on subsequent actions. This perspective allows for testing effects that consider group decisions on fine-grained temporal micro-mechanisms. Additionally, the model distinguishes between selection effects—where authors preferentially cite based on prior relational structures—and influence effects, in which network exposure shapes future citation behavior. In our case, we identify whether coherent author groups influence one another through shared citation practices (e.g., adopting the citation patterns of coauthors) or whether they extend their influence indirectly, such as by being cocited.

\hide{In our case, we were able to identify if the ``coherent groups'' can influence other members by promoting the authors to adopt the past citations of their past coauthors or to cite the papers of their coauthors, or if they might also expand the group by being cocited by third authors. }

While we have advanced the knowledge about socio-cognitive networks among researchers, several limitations need to be addressed. 
Our study's time frame was restrictive and only covered a short period ($2013-2015$). 
Further applications should incorporate extended periods. 
However, increasing the time window requires careful consideration of how reasonable the assumption is that authors are aware of all previous publications necessary to identify intercitations. 
Additionally, our study is only a case study. 
We believe that comparisons between disciplines and including interdisciplinary research areas can enhance the exploration of complex networks. 
Topics are a relevant area of research that should further explore the relevance of specializations. 
There are many possibilities for exploring this dimension. 
For instance, one can independently run a Large Language Model on the textual data to derive embeddings for the papers and then use this in our framework as covariates for the respective papers. 
 Alternatively, one may derive keywords from past papers to encode the areas of expertise of each author who contributed to the paper. 
One can use this information as exogenous covariates, such as the nationality of the authors (variable whether the authors are Chilean or not). Finally, moving beyond bibliometric data can also help explore the social dimension underpinning the scientific network. Further research should go beyond formal communication channels in science.

\bibliographystyle{chicago}
\bibliography{references}

\newpage

\begin{center}
{\LARGE\textbf{Supplementary Material:\\ 
Socio-cognitive Networks between Researchers}}
\end{center}

\setcounter{section}{0}

\setcounter{page}{1}

\appendix
\setcounter{equation}{0}

\renewcommand{\cftsecfont}{\mdseries}
\renewcommand{\cftsecpagefont}{\mdseries}
\setlength{\cftbeforesecskip}{0 pt} 
\renewcommand{\cftsecleader}{\cftdotfill{\cftdotsep}} 
\startcontents
\printcontents{ }{1}{}

\section{Related Models} 
\label{sec:related}

We define a probabilistic model for $\mathscr{W}(\mathscr{T}) = \{w; \, t(w) \in \mathscr{T}\}$ encompassing the published work of the astronomers in Chile between 2013 and 2015. 
To empirically assess the socio-cognitive mechanisms specified in Sections \ref{sec:theory} and \ref{sec:socio-cognitive}, this framework should allow the specification of structural covariates and accommodate high-order interactions.
Relational Event Models (REM) for dyadic interactions proposed by \citetsupp{butts_relational_2008} are commonly employed for timestamped data since they can use fine-grained temporal information without the need for aggregation, which would be necessary for alternative methods, such as Temporal Exponential Graph Models \citepsupp{robins_random_2001} and Stochastic Actor Oriented Models \citepsupp{Snijders_2001, snijders_introduction_2010}. 
Dyadic interactions cover settings where we are, e.g., interested in modeling the citations between works \citepsupp{filippi_mazzola_stochastic_2024} or co-authoring work \citepsupp{fritz_modelling_2023} but fail to capture higher-order interactions involving more than one sender and receiver for each event, which we call hyperevents. 
Adaptions of the framework based on latent variables  \citepsupp{rastelli_continuous_2023} or actor-oriented models \citepsupp{stadtfeld_dynamic_2017, Stadtfeld_Block_2017} are similarly not tailored towards hyperevents and are, in the former case, not able to incorporate theory-driven covariates derived from Section \ref{sec:theory}. 
Finally, bipartite extensions employed in \citepsupp{malangNetworksSocialInfluence2019} could be used for our modeling hyperevents. 
For the author model, the first mode would represent the authors, and the second mode would represent all published papers. 
Note, however, that the random outcome we model is no longer the set of authors but the separate decision of joining a particular project. 
Therefore, this bipartite representation is insufficient in representing joint decisions to write a joint paper.


\section{Interpretation of the Coefficients} 
\label{sec:interpretation}
One can interpret the coefficients $\theta$ and $\gamma$ in the same manner as coefficients for proportional hazard models. 
With $p \in \{1, \ldots, P\}$, take two possible author sets $A \subseteq \mathscr{A}$ and $A^\star\subseteq \mathscr{A}$ with $A \neq A^\star$ and let $\theta_p$ and $s_p(\mathscr{W}(t), A)$ refer to the $p$th coefficient and statistic of the author model in \eqref{eq:author}. 
If $s_k(\mathscr{W}(t), A) = s_k(\mathscr{W}(t), A^\star)$ for all $k \neq p$ and $s_p(\mathscr{W}(t), A) = s_p(\mathscr{W}(t), A^\star) + 1$, $\theta_p>0$ is the multiplicative change that we are more likely to observe the author set $A^\star$ than $A$. The same interpretation holds for $\gamma$ with the only difference that we compare two possible citation lists in a work while conditioning on the authors $A$ and size of cited works $C$.

We illustrate this interpretation of coefficients with the statistic \textsl{``Heterogeneity Chilean''}, which is the ratio of author-pairs $\{i,j\} \,\subset \, A$, such that $i$ is Chilean and $j$ is not ($\text{Chilean}(i) = 1$ and $\text{Chilean}(j) = 0$) and described in Section \ref{sec:attributes}. The values of this statistic theoretically range from zero (all authors in $A$ are Chilean, or none is Chilean) to one ($A$ contains exactly two authors, of which one is Chilean and the other is not). The coefficient of this statistic in the author model is $0.8$. This means that a set of authors $A$ with $s_{\text{Het. Chilean}}(\mathscr{W}(t), A)=1$ is predicted to be $\exp(0.8)\, =\, 2.23$ times more likely to be the set of coauthors of a published work than another set of authors $A\star$ with $s_{\text{Het. Chilean}}(\mathscr{W}(t), A^\star)$, all other statistics being equal. To provide another example, a set of authors $A$ with $s_{\text{Het. Chilean}}(\mathscr{W}(t), A)\, =\, 0.25$ is predicted to be $\exp(0.25\times 0.8)\, =\, 1.22$ times more likely to be the set of coauthors of a published work than a set of authors $A^\star$ with $s_{\text{Het. Chilean}}(\mathscr{W}(t), A^\star)\, =\, 0$.

\paragraph*{Transformation of statistics.}
We transform all endogenous statistics, that is, all statistics except \textsl{``Ratio Chilean''} and \textsl{``Heterogeneity Chilean''}, applying the square-root $x\mapsto\sqrt{x}$, as this typically leads to better model fit \citep{lerner2023relational_polyadic} by scaling down large values and hence attenuating skewness of statistics. Subsequently, we standardize statistics to mean zero (subtracting the mean value) and standard deviation one (dividing by the standard deviation). Division by the standard deviation scales parameters and standard errors in the opposite direction (i.\,e., multiplies them with the standard deviation) and hence does not affect parameter signs, z-values, or p-values. The standardization is motivated by considering one standard deviation a ``typical variation'' of statistics values among instances. A unit change in a statistic hence means a change by one standard deviation. Centering statistics to mean zero has no effect on estimated parameters -- already for the fact that the \texttt{coxph} function in the \texttt{survival} package centers covariates before estimation.

\section{Specification of Statistics} 
\label{sec:sufficient_statistics}
The statistics at time point $t \in \mathscr{T}$ for the set of authors $A \subseteq \mathscr{A}$ and set of cited works $C \subseteq \mathscr{W}(t)$, denoted by $s(\mathscr{W}(t), A)$ and $h(\mathscr{W}(t), C, A)$ for the author and citation model, respectively, are functions of $\mathscr{W}(t)$, which includes the entire bibliographic information up until but not including $t$. As in the notation introduced in the main manuscript, the static exogenous binary information on whether a particular astronomer in the data is Chilean is included in $\mathscr{W}(t)$. 
In Section \ref{sec:attributes}, we define several time-dependent functions, which we call ``\textsl{network attributes}'', summarizing past author-work, author-author, and work-work interactions. These network attributes act as building blocks for stating the statistics in our model, which we introduce consecutively (for the author model in Section \ref{sec:suff_stat_author_model} and the citation model in Section \ref{sec:suff_stat_citation_model}).

Most employed statistics are defined in \citet{lerner2024relational_coevolution}. To produce a self-contained manuscript and adapt to the notation used in this paper, we provide a complete list of formulae defining the full model specification employed in Section \ref{sec:results}. 

\subsection{Definition of Network Attributes}
\label{sec:attributes}

\paragraph*{Author-Work Interaction:}
The extent to which a set of authors $A \subseteq \mathscr{A}$ has cited a set of works $C \in \mathscr{W}(t)$ for $t \in \mathscr{T}$ in joint publications is given by: 
\beno
\label{eq:def_cite_a_w}
\text{cite}^{(a \rightarrow w)}_{\, t}(A,C)&=&\dsum_{l \in \mathscr{W}(t)} \mathbbm{1}(A \subseteq A(l)\wedge C\subseteq C(l))\enspace.
\ee
Let $\text{auth}_{\, t}^{(a \rightarrow w)}(i,h)$ be a binary attribute indicating if author $i \in \mathscr{A}$ was an author of work $h \in \mathscr{W}(t)$ 
\beno
\label{eq:def_coauthor}
\text{auth}_{\, t}^{(a \rightarrow w)}(i,h) &=& 
 \mathbbm{1}\big(i \in A(h)\big).
\ee

Another attribute that serves as a base to define several different types of statistics characterizing our model is called \textsl{subset repetition of order $(k,k^\star)$} with $k,k^\star \in \{0, 1, 2, ...\}$, not both being equal to zero. With this information, we quantify: (1) if a set of authors $A \subseteq \mathscr{A}$ already coauthored one or several publications (possibly together with others); (2) if a set of publications $C \in \mathscr{W}(t)$ published before $t \in \mathscr{T}$ has been co-cited (possibly within a larger list of references); (3) if a set of authors $A \subseteq \mathscr{A}$ have coauthored a publication citing a set of works $C \in \mathscr{W}(t)$. 
The two-dimensional order $(k,k^\star)$ with $k,k^\star \in \{0, 1, 2, ...\}$ of the subset repetition attribute evaluated for 
the set of authors $A \subseteq \mathscr{A}$ and set of publications $C \in \mathscr{W}(t)$ at $t \in \mathscr{T}$
relates to the sizes of the subsets of authors and works that are repeated: 
\beno
\label{eq:def_subrep}
\text{subrep}^{(k,k^\star)}_{\, t}(A,C) &=& \dsum_{(A^\star,C^\star)\, \in\, {\text{sub}(A, k)}\, \times \, {\text{sub}(C, k^\star)}}
\dfrac{\text{cite}^{(a \rightarrow w)}_{\, t}(A^\star,C^\star)}{{|A|\choose k}\cdot{|C|\choose k^\star}}\enspace,
\ee
where $\text{sub}(A, k)$ and $\text{sub}(C, k^\star)$ denote all possible subsets of actor set $A$ of size $k$ and all possible subsets of cited works $C$ of size $k^\star$, respectively.

\paragraph*{Author-Author Interaction:}
    For two authors $i,j\in\mathscr{A}$, past author-to-author citations are given by 
\beno
\label{eq:def_cite_a_a}
\text{cite}^{(a \rightarrow a)}_{\, t}(i,j)&=&\dsum_{l \in \mathscr{W}(t)} \mathbbm{1}\left(\text{auth}_{\, t}^{(a \rightarrow w)}(i,l)  \, \wedge\,   \left(\dsum_{m \in C(l)} \text{auth}_{\, t}^{(a \rightarrow w)}(j,m) \geq 1\right)\right)\enspace.
\ee

The network attribute $\text{p}_{\, t}(i)$ for $i \in \mathscr{A}$ and $t \in \mathscr{T}$ denotes the citation popularity of $i$, i.e., how much past work of $i$ is cited before $t$: 
\beno
\label{eq:def_cit_popularity}
\text{p}_{\, t}^{(a)}(i) &=& 
\dsum_{l \in \mathscr{W}(t)} \mathbbm{1}\left(\left(\dsum_{m \in C(l)} \text{auth}_{\, t}^{(a \rightarrow w)}(i,m)\right) \,\geq\, 1 \right).
\ee
Finally, we write 
\beno
\label{eq:def_coauth}
\text{coauth}_{\, t}^{(a)}(i,j) = \text{cite}^{(a \rightarrow w)}_{\, t}(\{i,j\},\emptyset)
\ee
for authors $i,j \in \mathscr{A}$ to indicate the number of coauthored papers of $i$ and $j$ before time $t \in \mathscr{T}$.

\paragraph*{Work-Work Interaction:}
    For the works $h, k \in\mathscr{W}(t)$ published before $t \in \mathscr{T}$, the work-to-work interactions are represented by 
\beno
\label{eq:def_cite_w_w}
\text{cite}^{(w \rightarrow w)}_{\, t}(k,h)&=&\mathbbm{1}(h\in C(k)),
\ee
being the indicator of whether work $h$ was cited in work $k$,

\subsection{Statistics of the Author Model}
\label{sec:suff_stat_author_model}
\begin{figure}[t!]
    \centering
    \begin{subfigure}[b]{0.3\textwidth}
        \includegraphics[clip, trim=0.5cm 1.5cm 0.5cm 0cm,width=\linewidth, page = 15]{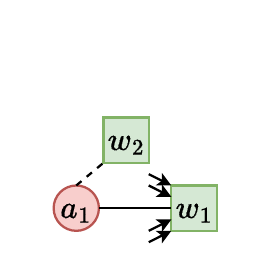}
        \caption{Citation Popularity\newline Author}
        \label{supfig:cit_pop}
    \end{subfigure}  
    \begin{subfigure}[b]{0.3\textwidth}
        \includegraphics[clip, trim=0.5cm 1.5cm 0.5cm 0cm,,width=\linewidth, page = 16]{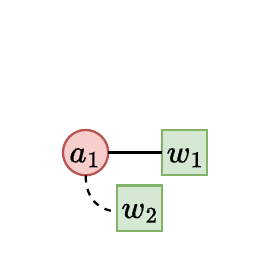}
        \caption{Publication Activity\newline~}
        \label{supfig:activity}
    \end{subfigure}  
    \begin{subfigure}[b]{0.3\textwidth}
        \includegraphics[clip, trim=0.5cm 1.5cm 0.5cm 0cm,,width=\linewidth, page = 17]{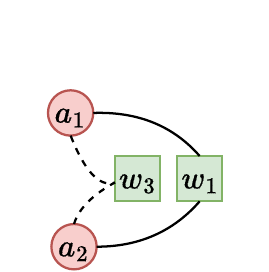}
        \caption{Coauthor-Pair \newline Repetition}
        \label{supfig:pair}
    \end{subfigure}  
    \begin{subfigure}[b]{0.3\textwidth}
        \includegraphics[clip, trim=0.5cm 1cm 0.5cm 0cm,width=\linewidth, page = 18]{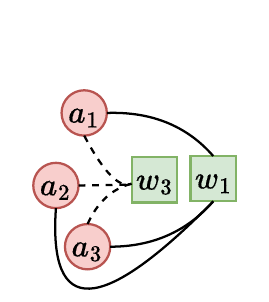}
        \caption{Coauthor-Triplet\newline Repetition}
        \label{supfig:tri}
    \end{subfigure}  
     \begin{subfigure}[b]{0.3\textwidth}
        \includegraphics[clip, trim=0.5cm 1cm 0.5cm 0cm,width=\linewidth, page = 19]{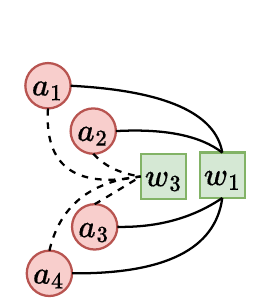}
        \caption{Coauthor-Quartet\newline Repetition}
        \label{supfig:quartet}
    \end{subfigure}  
    \begin{subfigure}[b]{0.3\textwidth}
        \includegraphics[clip, trim=0.5cm 1.5cm 0.5cm 0cm,width=\linewidth, page = 20]{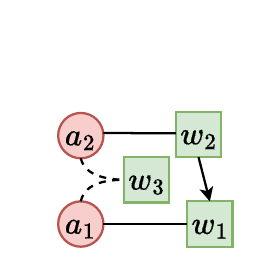}
        \caption{Collaborating with\newline Citing Author}
        \label{supfig:col_cit_auth}
    \end{subfigure} 
     \begin{subfigure}[b]{0.3\textwidth}
        \includegraphics[clip, trim=0.5cm 1.5cm 0.5cm 0cm,width=\linewidth, page = 21]{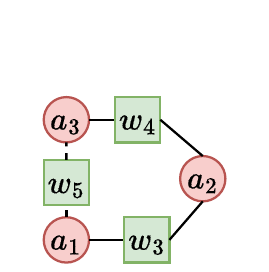}
        \caption{Closure by Coauthor}
        \label{supfig:clos_by_auth}
    \end{subfigure}
         \begin{subfigure}[b]{0.3\textwidth}
        \includegraphics[clip, trim=0.5cm 1.5cm 0.5cm 0cm,width=\linewidth, page = 22]{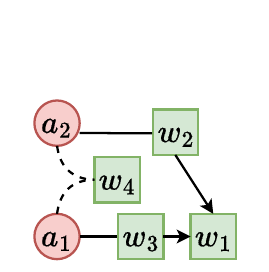}
        \caption{Closure by Work}
        \label{supfig:clos_by_work}
    \end{subfigure}
    \caption{Statistics included in the Author Model: Illustrative illustration of the relationships between authors and their works, as well as citation dynamics among works. 
    Red circles represent authors, and green rectangles denote works. Solid lines indicate realized authorship connections between authors and their works, while dashed lines represent potential authorship relationships. Solid arrows show actual citation relationships between works}
    \label{supfig:statistcs_author}
\end{figure}

The vector of statistics in the author model $s(\mathscr{W}(t), A)\in \mathbb{R}^P$, encompasses $P \in \{1, 2, \ldots\}$ separate terms, each capturing different facets how the author team $A \subseteq \mathscr{A}$ is determined at time $t \in \mathscr{T}$ given the bibliographic information from the past, which is denoted by $\mathscr{W}(t)$. 
Next, we describe each entry of this vector separately. 
We denote each entry of $s(\mathscr{W}(t), A)$ by its name. For instance, if the statistic is called \textsl{``text''}, the statistic is denoted by $s_{\text{text}}(\mathscr{W}(t), A)$ with corresponding coefficient $\theta_{\text{text}}$.
In some cases, we shorten a statistic's name for better readability. 
All statistics are visualized in Figure \ref{supfig:statistcs_author} and are given by: 
\beno 
s(\mathscr{W}(t), A) &=& \Big(s_{\text{Ratio Chilean}}(\mathscr{W}(t), A), s_{\text{Het. Chilean}}(\mathscr{W}(t), A), \\ &~& ~~
f(s_{\text{Citation Pop. Author}}(\mathscr{W}(t), A)),f(s_{\text{Coauthor-Pair Rep.}}(\mathscr{W}(t), A)), \\ &~&~~
f(s_{\text{Coauthor-Triplet Rep.}}(\mathscr{W}(t), A)),f(s_{\text{Coauthor-Quartet Rep.}}(\mathscr{W}(t), A)), \\ &~&~~
f(s_{\text{Coll. with Citing Author}}(\mathscr{W}(t), A)),f(s_{\text{Clos. by Coauthor}}(\mathscr{W}(t), A)), \\ &~&~~
f(s_{\text{Clos. by Work}}(\mathscr{W}(t), A))\Big),
\ee
where $f: \mathbb{R} \mapsto \mathbb{R}$ is the transformation function (square root and normalization) described in Section \ref{sec:interpretation}.

\paragraph*{Exogenous Information on Authors}

We define the ratio of Chilean scientists in a set of authors $A\subseteq\mathcal{A}$ via the covariate
\beno 
s_{\text{Ratio Chilean}}(\mathscr{W}(t), A) &=&\dsum_{i\,\in\, A}
\dfrac{\text{Chilean}(i)}{|A|}\enspace,
\ee
where $\text{Chilean}(i)$ is the binary indicator that is one if $i$ is Chilean and zero otherwise. If $\theta_{\text{\, Ratio Chilean}}>0$, the model suggests that Chileans are more likely to be included in the set of coauthors or, from another point of view, that Chileans publish at a higher rate than non-Chileans in our data set. 

We define the heterogeneity of a group of scientists $A\subseteq\mathcal{A}$ with respect to Chilean nationality via the covariate
\beno 
s_{\text{Het. Chilean}}(\mathscr{W}(t), A) &=&\dsum_{\{i,j\}\,\in\, \text{sub}(A, 2)}
\dfrac{|\text{Chilean}(i)-\text{Chilean}(j)|}{\binom{|A|}{2}}\enspace,
\ee
If $\theta_{\text{\, Het. Chilean}}>0$, the model suggests that groups of authors tend to be more diverse with respect to Chilean nationality than expected by random selection of authors, that is, that groups of authors tend to mix Chileans with Non-Chileans.

\paragraph*{Citation Popularity of Author (Figure \ref{supfig:cit_pop})}
We define the average citation popularity of a set of authors $A\subseteq\mathcal{A}$ via the covariate
\beno 
s_{\text{Citation Pop. Author}}(\mathscr{W}(t), A) &=&\dsum_{i\in A}
\frac{\text{p}^{(a)}_{\, t}(i)}{|A|}\enspace.
\ee
If $\theta_{\text{\, Citation Pop. Author}}>0$, the model suggests that authors whose works have been cited more often in the past are more likely to publish future papers.

\paragraph*{Publication Activity by Groups of Authors (Figures \ref{supfig:activity}, \ref{supfig:pair}, \ref{supfig:tri}, and \ref{supfig:quartet})}
For a set of authors $A\subseteq\mathscr{A}$, the average number of prior joint work is given by the statistic
\beno 
s_{\text{Publication Act.}}(\mathscr{W}(t), A)&=& \text{subrep}^{(1,0)}_{\, t}(A,\emptyset),
\ee
which can be comprehended as a measure of the average past publication activity of authors in $A$. If $\theta_{\text{Publication Activity}}>0$, then the model suggests that authors who have published more in the past publish at a higher rate in the future.

Previous collaboration among pairs of authors is captured by 
\beno 
s_{\text{Coauthor-Pair Rep.}}(\mathscr{W}(t), A)&=& \text{subrep}^{(2,0)}_{\, t}(A,\emptyset),
\ee
which averages the number of coauthored papers over all unordered pairs of authors in $A$. 
The statistics $s_{\text{Coathor-Triple Repetition}}(\mathscr{W}(t), A)$ and $s_{\text{Coathor-Quartet Repetition}}(\mathscr{W}(t), A)$ are defined along the same lines by using subset repetition of order $(3,0)$ and $(4,0)$. If $\theta_{\text{Coauthor-Pair Repetition}}>0$, then the model suggests that pairs of authors who have co-authored more papers in the past are more likely to be coauthors in the future. If $\theta_{\text{Coauthor-Triple Repetition}}>0$, then the model suggests that triples of authors who have jointly co-authored more papers in the past are more likely to be coauthors in the future -- on top of what is possibly explained by Coauthor-Pair Repetition. If $\theta_{\text{Coauthor-Quartet Repetition}}>0$, then the model suggests that sets of four authors who have jointly co-authored more papers in the past are more likely to be coauthors in the future -- on top of what is possibly explained by Repetition of Coauthor-Pairs or Triples. 

\paragraph*{Collaborate with Citing Author (Figure \ref{supfig:col_cit_auth})}
The tendency of authors to coauthor works with those who cited the authors previous work is captured by a statistic measuring the past citation density within a set of authors $A\subseteq\mathscr{A}$:
\beno 
s_{\text{Coll. with Citing Author}}(\mathscr{W}(t), A)&=& \dsum_{\{i,j\}\in {\text{sub}(A,2)}}
\dfrac{\text{cite}^{(a \rightarrow a)}_{\, t}(i,j) + \text{cite}^{(a \rightarrow a)}_{\, t}(j,i)}{2\binom{|A|}{2}}\enspace.
\ee
If $\theta_{\text{Coll. with Cit. Author}}>0$, then the model suggests that two authors $i,j$ are more likely to coauthor a paper in the future if $i$ has cited papers of $j$ in the past, and/or if $j$ has cited papers of $i$ in the past.

\paragraph*{Closure by Author (Figure \ref{supfig:clos_by_auth})} Using the definition of $\text{coauth}_{\, t}^{(a)}(i,j)$ from \eqref{eq:def_coauth}, the extent of authors in $A\subseteq\mathscr{A}$ to coauthor works with the same \textsl{``third''} author is captured by:  
\be
\label{eq:def_closure_by_author}
s_{\text{Clos. by Coauthor}}(\mathscr{W}(t), A)&=& \dsum_{\{i,j\}\, \in \, {\text{sub}(A,2)}} \, \dsum_{k \in \mathscr{A}: k\, \neq\, i,\, j}
\dfrac{\min\{\text{coauth}_{\, t}^{(a)}(i,k),\text{coauth}^{(a)}_{\, t}(j,k)\}}{{|A|\choose 2}}\enspace,
\ee
where $\min\{X\} \in \mathbb{R}$ denotes the minimum value in set $X \subset \mathbb{R}$. The \textsl{``third''} author in \eqref{eq:def_closure_by_author} is $k \in \mathscr{A}$, which has to be different from the authors of $i,j \in A$.
If $\theta_{\text{Clos. By Auth.}}>0$, then the model suggests that two authors $i,j$ are more likely to coauthor a publication in the future if there is one (or more) ``third'' author $k$ such that $i$ has coauthored with $k$ and $j$ has coauthored (possibly different papers) with $k$.

\paragraph*{Closure by Citing same Work (Figure \ref{supfig:clos_by_work})}
A related covariate captures how much authors in $A\subseteq\mathscr{A}$ have cited the same works in the past:
\beno
&~& s_{\text{Clos. by Work}}(\mathscr{W}(t), A) \\ &=& \dsum_{\{i,j\}\, \in \, {\text{sub}(A,2)}} \, \dsum_{l \in \mathscr{W}(t)}
\dfrac{\min\{\text{cite}^{(a \rightarrow w)}_{\, t}(\{i\},\{l\}),\text{cite}^{(a \rightarrow w)}_{\, t}(\{j\},\{l\})\}}{{|A|\choose 2}}\enspace.
\ee
If $\theta_{\text{Clos. By Cit. same Work}}>0$, then the model suggests that two authors $i,j$ are more likely to coauthor a publication in the future if $i$ and $j$ have cited the same work(s) $l$ in the past (possibly when publishing different publications).

\subsection{Statistics of the Citation Model}
\label{sec:suff_stat_citation_model}

\begin{figure}[t!]
    \centering
    \begin{subfigure}[b]{0.3\textwidth}
        \includegraphics[clip, trim=0.5cm 1cm 0.5cm 0cm,width=\linewidth, page = 1]{Figures/all_Stats_no_caption.pdf}
        \caption{Citation Popularity of Work}
        \label{supfig:cit_pop_work}
    \end{subfigure}  
    \begin{subfigure}[b]{0.3\textwidth}
        \includegraphics[clip, trim=0.5cm 1cm 0.5cm 0cm,width=\linewidth, page = 2]{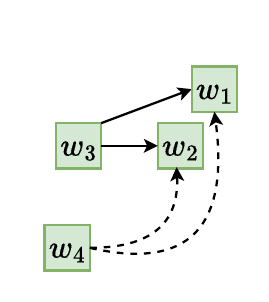}
        \caption{Cocitation Popularity: Pair}
        \label{supfig:cocite_pop_pair}
    \end{subfigure}  
    \begin{subfigure}[b]{0.3\textwidth}
        \includegraphics[clip, trim=0.5cm 1cm 0.5cm 0cm,width=\linewidth, page = 3]{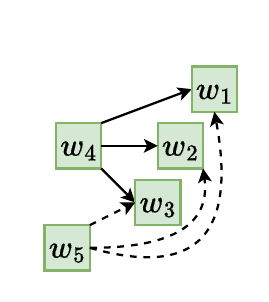}
        \caption{Cocitation Popularity: Triplet}
        \label{supfig:cocite_pop_tri}
    \end{subfigure}  
    \begin{subfigure}[b]{0.3\textwidth}
        \includegraphics[clip, trim=0.5cm 1cm 0.5cm 0cm,width=\linewidth, page = 4]{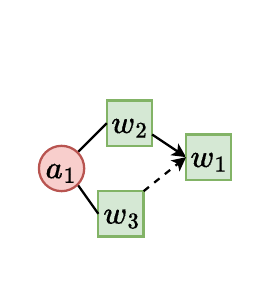}
        \caption{Citation Repetition\newline ~}
        \label{supfig:citation_rep}
    \end{subfigure}  
     \begin{subfigure}[b]{0.3\textwidth}
        \includegraphics[clip, trim=0.5cm 1cm 0.5cm 0cm,width=\linewidth, page = 5]{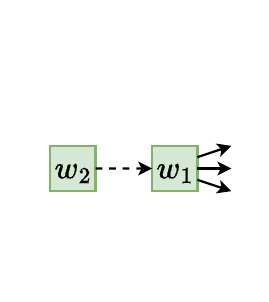}
        \caption{Outdegree Popularity\newline ~}
        \label{supfig:outdegree_pop}
    \end{subfigure}  
    \begin{subfigure}[b]{0.3\textwidth}
        \includegraphics[clip, trim=0.5cm 1cm 0.5cm 0cm,width=\linewidth, page = 6]{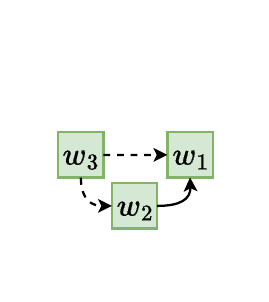}
        \caption{Cite Work \newline and its Citations}
        \label{supfig:cite_work_and_ref}
    \end{subfigure} 
    \caption{Statistics included in the Citation Model: Illustrative illustration of the relationships between authors and their works, as well as citation dynamics among works. 
    Red circles represent authors, and green rectangles denote works. Solid lines indicate authorship connections between authors and their works. Solid arrows show actual citation relationships between works, while dashed arrows represent potential citation relationships. }
    \label{supfig:statistcs1}
\end{figure}

\begin{figure}[t!]
    \centering
    \begin{subfigure}[b]{0.3\textwidth}
        \includegraphics[clip, trim=0.5cm 1cm 0.5cm 0cm,width=\linewidth, page = 7]{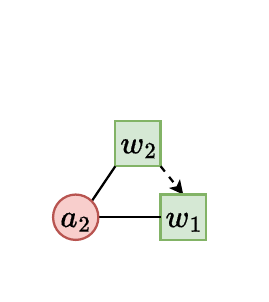}
        \caption{Self Citation}
        \label{supfig:self_citation}
    \end{subfigure} 
         \begin{subfigure}[b]{0.3\textwidth}
        \includegraphics[clip, trim=0.5cm 1cm 0.5cm 0cm,width=\linewidth, page = 8]{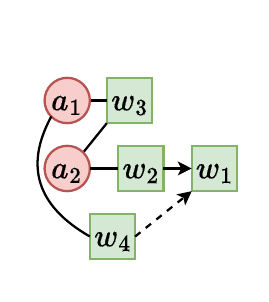}
        \caption{Adopt Citation of Coauthor}
        \label{supfig:adopt_cit_of_coauth}
    \end{subfigure} 
         \begin{subfigure}[b]{0.3\textwidth}
        \includegraphics[clip, trim=0.5cm 1cm 0.5cm 0cm,width=\linewidth, page = 9]{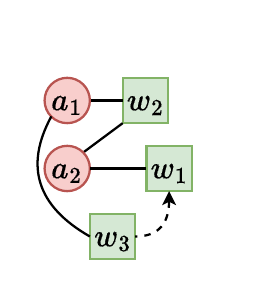}
        \caption{Cite Work of Coauthor}
        \label{supfig:cite_work_of_coauthor}
    \end{subfigure} 
    
    \begin{subfigure}[b]{0.3\textwidth}
        \includegraphics[clip, trim=0.5cm 1cm 0.5cm 0cm,width=\linewidth, page = 10]{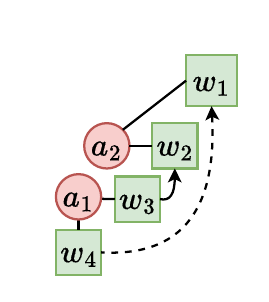}
        \caption{Author cites Author Repetition}
        \label{supfig:auth_cite_auth_rep}
    \end{subfigure}  
    \begin{subfigure}[b]{0.3\textwidth}
        \includegraphics[clip, trim=0.5cm 1cm 0.5cm 0cm,width=\linewidth, page = 11]{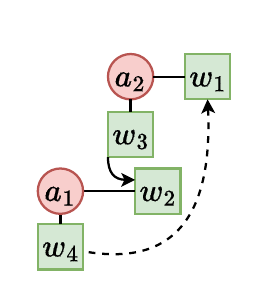}
        \caption{Author cites Author Reciprocity}
        \label{supfig:auth_cite_auth_rec}
    \end{subfigure}  
    \begin{subfigure}[b]{0.3\textwidth}
        \includegraphics[clip, trim=0.5cm 1cm 0.5cm 0cm,width=\linewidth, page = 12]{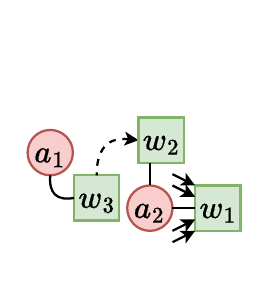}
        \caption{Cite much Cited Author\newline ~ }
        \label{supfig:cite_much_cited_auth}
    \end{subfigure}  
    \begin{subfigure}[b]{0.3\textwidth}
        \includegraphics[clip, trim=0.5cm 1cm 0.5cm 0cm,width=\linewidth, page = 13]{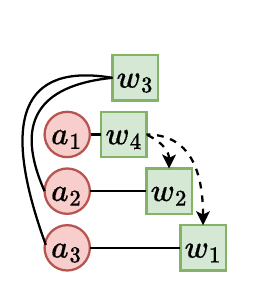}
        \caption{Cocite Coauthor Pairs}
        \label{supfig:cocite_coauthor_pairs}
    \end{subfigure}  
     \begin{subfigure}[b]{0.3\textwidth}
        \includegraphics[clip, trim=0.5cm 1cm 0.5cm 0cm, width=\linewidth, page = 14]{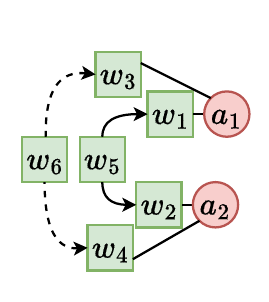}
        \caption{Author Cocitation}
        \label{supfig:cocite_cocited_authors}
    \end{subfigure}  
    \caption{Statistics included in the Citation Model (continued):  Illustrative illustration of the relationships between authors and their works, as well as citation dynamics among works. 
    Red circles represent authors, and green rectangles denote works. Solid lines indicate authorship connections between authors and their works. Solid arrows show actual citation relationships between works, while dashed arrows represent potential citation relationships.}
    \label{supfig:statistcs2}
\end{figure}

The vector of sufficient statistics of the citation model $h(\mathscr{W}(t), C, A)\in \mathbb{R}^Q$ encompasses $Q \in \{1, 2, \ldots\}$ separate terms, each capturing different facets how the set of citations $C \subseteq \mathscr{W}(t)$ is determined at time $t \in \mathscr{T}$ given the bibliographic information from the past, which is denoted by $\mathscr{W}(t)$, and the author team $A \subseteq \mathscr{A}$. 
Similar to Section \ref{sec:suff_stat_author_model}, we describe each entry of this vector separately. 
We denote each entry of $h(\mathscr{W}(t), C, A)$ by its name. For instance, if the statistic is called \textsl{``text''}, the statistic is denoted by $h_{\text{text}}(\mathscr{W}(t), C, A)$ with corresponding coefficient $\gamma_{\text{text}}$.
In some cases, we shorten a statistic's name for better readability.
All statistics are visualized in Figures \ref{supfig:statistcs1} and \ref{supfig:statistcs2} and are given by: 
\beno 
h(\mathscr{W}(t), C, A)&=& \Big(
f(h_{\text{\,Cit. Pop. of Work}}(\mathscr{W}(t), C, A)),\, 
f(h_{\text{\,Cocitation Pop.: Pair}}(\mathscr{W}(t), C, A)), \\ &~&~~
f(h_{\text{\,Cocitation Pop.: Tri.}}(\mathscr{W}(t), C, A)),\, 
h_{\text{\,Citation Rep.}}(\mathscr{W}(t), C, A)), \\ &~&~~
f(h_{\text{\,Outdegree Pop.}}(\mathscr{W}(t), C, A)),\,
f(h_{\text{\,Cite Work and its Cit.}}(\mathscr{W}(t), C, A)), \\ &~&~~
f(h_{\text{\,Self-Citation}}(\mathscr{W}(t), C, A)), \,
f(h_{\text{\,Adopt Cit. of Coauth.}}(\mathscr{W}(t), C, A)),  \\ &~&~~
f(h_{\text{\,Cite Work of Coauthor}}(\mathscr{W}(t), C, A)),\,
f(h_{\text{\,Author Cit. Author Rep.}}(\mathscr{W}(t), C, A)),  \\ &~&~~
f(h_{\text{\,Author Cit. Author Rec.}}(\mathscr{W}(t), C, A)), \,
f(h_{\text{\,Cite much Cited Authors}}(\mathscr{W}(t), C, A)),  \\ &~&~~
f(h_{\text{\,Cocite Coauthor Pairs}}(\mathscr{W}(t), C, A)), \,
f(h_{\text{\,Author Cocitation}}(\mathscr{W}(t), C, A))
\Big),
\ee
where $f: \mathbb{R} \mapsto \mathbb{R}$ is the transformation function (square root and normalization) described in Section \ref{sec:interpretation}.

\paragraph*{Citation and Co-citation Popularity (Figures \ref{supfig:cit_pop_work}, \ref{supfig:cocite_pop_pair}, and \ref{supfig:cocite_pop_tri})}
For a set of works $C\subseteq\mathscr{W}(t)$ that can be cited by author set $A\subseteq \mathscr{A}$ at time $t \in \mathscr{T}$, the average number of past citations is captured by the statistic \textsl{``Citation Popularity of Work''} 
\beno 
h_{\text{\,Cit. Pop. of Work}}(\mathscr{W}(t), C, A)&=&  \text{subrep}^{(0,1)}_{\, t}(\emptyset, C).
\ee 
If $\gamma_{\text{\,Cit. Pop. of Work}} > 0$, a ``rich get richer'' effect is suggested such that academic works with a higher number of citations until time $t$ are also more likely to be cited in the future. 
Co-citations of pairs and triples of publications are captured by
\beno 
h_{\text{\,Cocitation Pop.: Pair}}(\mathscr{W}(t), C, A)&=&  \text{subrep}^{(0,2)}_{\, t}(\emptyset, C) \\
h_{\text{\,Cocitation Pop.: Tri.}}(\mathscr{W}(t), C, A)&=&  \text{subrep}^{(0,3)}_{\, t}(\emptyset, C).
\ee
Positive parameters for these covariates suggest that pairs (or triples) of publications that were jointly cited before time $t$ are more likely to be co-cited again in the future.

\paragraph*{Citation Repetition (Figure \ref{supfig:citation_rep})}
To check if some authors in $A \subseteq \mathscr{A}$ repeatedly cite the same work in $C \subseteq \mathscr{W}(t)$, we incorporate the following statistic:
\beno 
h_{\text{\,Citation Rep.}}(\mathscr{W}(t), C, A)&=&  \text{subrep}^{(1,1)}_{\, t}(A, C).
\ee
For $\gamma_{\text{\,Citation Rep.}} > 0$, the model implies that works previously cited by authors in $A$ before time $t$ are more likely to be cited again by authors in $A$.

\paragraph*{Outdegree Popularity (Figure \ref{supfig:outdegree_pop})}
The average length of reference lists of a set of works $C\subseteq\mathscr{W}(t)$ published before time $t \in \mathscr{T}$ is given by 
\beno 
h_{\text{\,Outdegree Pop.}}(\mathscr{W}(t), C, A)&=&  \dfrac{1}{|C|} \dsum_{l \in C} |l|.
\ee
If $\gamma_{\text{\,Outdegree Pop.}} > 0$, we can interpret the result as a tendency to prefer citing works that, in turn, cite many other publications. 
 
\paragraph*{Cite Work and its Citations (Figure \ref{supfig:cite_work_and_ref})}
The tendency to adopt (some of) the references of a cited work is represented by the past citation density within a set of cited works $C\subseteq\mathscr{W}(t)$ published before time $t \in \mathscr{T}$:
\beno 
h_{\text{\,Cite Work and its Cit.}}(\mathscr{W}(t), C, A)&=&  \dsum_{\{h,k\}\,\in\, \text{sub}(C,2)} \dfrac{\text{cite}^{(w \rightarrow w)}_{\, t}(h,k) + \text{cite}^{(w \rightarrow w)}_{\, t}(k,h)}{\binom{|C|}{2}}.
\ee
If $\gamma_{\text{\,Cite Work and its Cit.}} > 0$, the model suggest that two publications $h$ and $k$ are more likely to be cocited in a future publication if $h$ has cited $k$ or if $k$ has cited $h$.

\paragraph*{Self-Citation (Figure \ref{supfig:self_citation})}
To understand whether authors tend to cite their own past work we define the statistic \textsl{``Self-Citation''} being the density of the two-mode subgraph connecting a set of authors $A\subseteq\mathscr{A}$ and a set of cited works $C\subseteq\mathscr{W}(t)$:
\beno 
h_{\text{\,Self-Citation}}(\mathscr{W}(t), C, A)&=&  \dsum_{i \in A,\;k\in C} \dfrac{\text{auth}_{\, t}^{(a \rightarrow w)}(i,k)}{|A| \, \cdot \, |C|}.
\ee
If $\gamma_{\text{\,Self-Citation}} > 0$, the model suggest that a publication $k$ is more likely to be cited by a publication having $i$ among its authors, if $i$ is an author of $k$.

\paragraph*{Adopt Citation of Coauthor (Figure \ref{supfig:adopt_cit_of_coauth})}
The amount to which authors $A \subseteq \mathscr{A}$ cite works that have been cited before $t \in \mathscr{T}$ by their coauthors is captured by the following statistic:
\beno 
h_{\text{\,Adopt Cit. of Coauth.}}(\mathscr{W}(t), C, A) &=& \dsum_{i\, \in\, A}\, \dsum_{j \,\neq\, i}\,\dsum_{l \,\in \, C}
\dfrac{\min\{\text{coauth}^{(a)}_{\, t}(i,j),\text{cite}^{(a \rightarrow w)}_{\, t}(\{j\},\{l\})\}}{|A|\, \cdot \, |C|}\enspace.
\ee
If $\gamma_{\text{\,Adopt Cit. of Coauth}} > 0$, the model suggest that a publication $l$ is more likely to be cited by a publication having $i$ among its authors, if there is one, or several, past coauthors $j$ of $i$ who has cited $l$ in their prior work.

\paragraph*{Cite Work of Coauthor (Figure \ref{supfig:cite_work_of_coauthor})}
Similarly, the propensity of authors to cite work that has been published by their past coauthors is measured by 
\beno 
h_{\text{\,Cite Work of Coauthor}}(\mathscr{W}(t), C, A) &=& \dsum_{i\, \in\, A}\, \dsum_{j \,\neq\, i}\,\dsum_{l \,\in \, C}
\dfrac{\min\left\{\text{coauth}^{(a)}_{\, t}(i,j),\text{auth}_{\, t}^{(a \rightarrow w)}(j,l)\right\}}{|A|\, \cdot \, |C|}\enspace.
\ee
If $\gamma_{\text{\,Cite Work of Coauthor}} > 0$, the model suggest that a publication $l$ is more likely to be cited by a publication having $i$ among its authors, if there is one (or several) past coauthor $j$ who is an author of $l$.

\paragraph*{Author Cites Author Repetition (Figure \ref{supfig:auth_cite_auth_rep})}
The tendency of authors to repeatedly cite the work of the same authors is measured by 
\beno 
h_{\text{\,Author Cit. Author Rep.}}(\mathscr{W}(t), C, A) &=& \dsum_{i\, \in\, A}\, \dsum_{j \,\neq\, i}\,\dsum_{l \,\in \, C}
\dfrac{\min\left\{\text{cite}^{(a \rightarrow a)}_{\, t}(i,j),\text{auth}_{\, t}^{(a \rightarrow w)}(j,l)\right\}}{|A|\, \cdot \, |C|}\enspace.
\ee
If $\gamma_{\text{\,Author Cit. Author Rep}} > 0$, the model suggest that a publication $l$ is more likely to be cited by a publication having $i$ among its authors, if there is one (or several) author $j$ of $l$, such that $i$ has already cited one (or several) past work of $j$ (note that the past work of $j$ that has been cited by $i$ may be different from $l$).

\paragraph*{Author Cites Author Reciprocation (Figure \ref{supfig:auth_cite_auth_rec})}
The statistic \textsl{``Author Cites Author Reciprocation''} captures the extend to which authors cite works of other authors who have previously cited their work: 
\beno 
h_{\text{\,Author Cit. Author Rec.}}(\mathscr{W}(t), C, A) &=& \dsum_{i\, \in\, A}\, \dsum_{j \,\neq\, i}\,\dsum_{l \,\in \, C}
\dfrac{\min\left\{\text{cite}^{(a \rightarrow a)}_{\, t}(j,i),\text{auth}_{\, t}^{(a \rightarrow w)}(j,l)\right\}}{|A|\, \cdot \, |C|}\enspace.
\ee
If $\gamma_{\text{\,Author Cit. Author Rec}} > 0$, the model suggest that a publication $l$ is more likely to be cited by a publication having $i$ among its authors, if there is one (or several) author $j$ of $l$, such that $j$ has already cited one (or several) past work of $i$. The difference to the author cites author repetition is that here the direction of past citations among $i$ and $j$ are reversed.

\paragraph*{Cite much Cited Authors (Figure \ref{supfig:cite_much_cited_auth})} The tendency of citing work of authors whose (potentially other) work has received many citations before time $t \in \mathscr{T}$ is 
 captured in our model via the following statistic:
\beno
h_{\text{\,Cite much Cited Authors}}(\mathscr{W}(t), C, A)&=&\dsum_{l\, \in\, C}
\dfrac{\max\left\{\text{p}_{\, t}^{(a)}(i)\colon\; i\in A(l)\right\}}{|C|}\enspace.
\ee
For each of the works $l\in C$, we use the maximum citation popularity of any author $i$ of $l$. Recall that the citation popularity of an author $i$ at time $t$ is the cumulative number of citations that any of $i$'s publication received strictly before $t$. This maximum citation popularity is then averaged over all works in the possible list of citations $C$. 
If $\gamma_{\text{\,Cite much Cited Authors}} > 0$, the model suggests that a paper $l$ is more likely to be cited if it has a scientist $i$ with high citation popularity among its authors.

\paragraph*{Cocite Coauthor Pairs (Figure \ref{supfig:cocite_coauthor_pairs})} The tendency to cocite pairs of publications that have been written by former coauthors is measured by the following statistics:
\beno
&~&h_{\text{\,Cocite Coauthor Pairs}}(\mathscr{W}(t), C, A)\\ 
&=&\dsum_{\{k,l\}\, \in\, \binom{C}{2}}
\frac{\mathbbm{1}\left\{\exists\, i,j \in \mathscr{A}\,\colon\; (\text{coauth}_{\, t}^{(a)}(i,j)>0) \wedge \text{auth}_{\, t}^{(a \rightarrow w)}(i,k)\wedge \text{auth}_{\, t}^{(a \rightarrow w)}(j,l)\right\}}
{\binom{|C|}{2}}\enspace.
\ee
The statistic computes the fraction of all pairs of works $k,l$ in C in which an author $i$ of $k$ and an author $j$ of $l$ have coauthored at least one publication before $t$.
If $\gamma_{\text{\,Cocite Coauthor Pairs}} > 0$, the model suggests that two papers $k,l$ are more likely to be cocited if there are former coauthors $i,j$ such that $i$ is an author of $k$ and $j$ is an author of $l$. Note that the publication(s) coauthored by $i$ and $j$ may be different from $k$ and different from $l$.

\paragraph*{Author Cocitation (Figure \ref{supfig:cocite_cocited_authors})} The tendency to cocite pairs of publications that have been written by formerly cocited authors is measured by the following statistics.
\beno
&~& h_{\text{\,Author Cocitation}}(\mathscr{W}(t), C, A) \\ 
&=&\dsum_{\{k,l\}\, \in\, \binom{C}{2}}
\frac{\mathbbm{1}\left\{\exists\, i,j \in \mathscr{A}\,\colon\; (\text{cocite}_{\, t}^{(a)}(i,j)>0) \wedge \text{auth}_{\, t}^{(a \rightarrow w)}(i,k)\wedge \text{auth}_{\, t}^{(a \rightarrow w)}(j,l)\right\}}
{\binom{|C|}{2}}\enspace.
\ee
where $\text{cocite}_{\, t}^{(a)}(i,j)$ denotes the count of publications published before $t$ that cite at least one publication of authors $i$ and at least one publication of authors $j$ with $i\neq j$ and $i,j \in \mathscr{A}$. The statistic computes the fraction of all pairs of works $k,l$ in $C$, such that there is an author $i$ of $k$ and there is an author $j$ of $l$ that have been cocited by at least one publication before $t$.
If $\gamma_{\text{\,Author Cocitation}} > 0$, the model suggests that two publications $k,l$ are more likely to be cocited if there is an author $i$ of $k$ and there is an author $j$ of $l$ such that $i$ and $j$ have been cocited before by some work. 

\section{Estimation} 
\label{sec:estimation}
Estimation of the unknown parameters $\theta$ and $\gamma$ is carried out by separately maximizing a case-control approximation of the partial likelihood arising from models \eqref{eq:author} and \eqref{eq:citation} (for further information on this, we refer to \citealpsupp{lerner2024relational_coevolution}). For estimating the authors model, we sample for every observed set of authors $A$ up to 30,000 randomly selected alternative sets of authors $A^\star\neq A$ (\textsl{``non-events''} or \textsl{``controls''}) with $|A^\star|=|A|$.  For estimating the citation model, we sample for every observed set of references $C$ up to 10,000 randomly selected alternative sets of works $C^\star\neq C$ (\textsl{``non-events''} or \textsl{``controls''}) with $|C^\star|=|C|$.
If the risk set size is smaller than 30,000 or 10,000 in the respective model, we use the entire risk set. (This is especially likely to happen in the author model for small sets of authors.)

We compute statistics of all events and sampled controls with the \texttt{eventnet} software \citepsupp{lerner2023relational_polyadic} and estimate parameters with the \texttt{coxph} function of the R-package \texttt{survival} \citepsupp{therneau2024survival-package}, using robust estimation \citepsupp{Therneau_2000}. 
This is recommended for these types of models in cases where it cannot be guaranteed that the specified statistics capture all dependence on the past; see \citetsupp{aalen_survival_2008} and the discussion given in \citetsupp{lerner2024relational_coevolution}.

\bibliographystylesupp{chicago}
\bibliographysupp{references}

\end{document}